\documentclass[]{aa}
\pdfoutput=1
\usepackage{geometry} 
\usepackage{graphicx}
\usepackage[varg]{txfonts}
\usepackage{upgreek}
\usepackage{soul}
\bibpunct{(}{)}{;}{a}{}{,} 

\newcommand{\radu}{rad m$^{-2}$}

\newcommand{\inv}{$^{-1}$}

\usepackage{microtype}
\title{Faraday Tomography of the Local Interstellar Medium with LOFAR: Galactic Foregrounds Towards IC342}
\titlerunning{Faraday Tomography of the Local ISM}
\author{C.L.~Van Eck\inst{1} \and 
M.~Haverkorn\inst{1} \and
M.I.R.~Alves\inst{2} \and
R.~Beck\inst{3} \and
A.G.~de Bruyn\inst{4,5} \and
T.~En{\ss}lin\inst{6,7} \and
J.S.~Farnes\inst{1} \and
K.~Ferri\`{e}re\inst{2} \and
G.~Heald\inst{8,5} \and
C.~Horellou\inst{9} \and
A.~Horneffer\inst{3} \and 
M.~Iacobelli\inst{4} \and
V.~Jeli\'{c}\inst{10, 4} \and
I.~Mart\'{i}-Vidal\inst{9} \and
D.D.~Mulcahy\inst{11} \and
W.~Reich\inst{3} \and
H.J.A.~R\"{o}ttgering\inst{12} \and
A.M.M~Scaife\inst{11} \and
D.H.F.M.~Schnitzeler\inst{3} \and
C.~Sobey\inst{13,8,4} \and
S.S.~Sridhar\inst{5,4} }
\authorrunning{C. Van Eck. et al.}
\institute{Department of Astrophysics/IMAPP, Radboud University, PO Box 9010, NL-6500 GL Nijmegen, the Netherlands;\thanks{The Faraday depth cube is available in electronic form at the CDS via anonymous ftp to cdsarc.u-strasbg.fr (130.79.128.5) or via http://cdsweb.u-strasbg.fr/cgi-bin/qcat?J/A+A/.} 
\email{c.vaneck@astro.ru.nl} 
\and 
IRAP, Universit\'{e} de Toulouse, CNRS, 9 avenue du Colonel Roche, BP 44346, 31028, Toulouse Cedex 4, France \and 
Max-Planck-Institut f\"{u}r Radioastronomie, Auf dem H\"{u}gel 69, 53121 Bonn, Germany \and 
ASTRON, the Netherlands Institute for Radio Astronomy, Postbus 2, 7990 AA Dwingeloo, The Netherlands \and 
Kapteyn Astronomical Institute, PO Box 800, 9700 AV Groningen, The Netherlands \and 
Max Planck Institute for Astrophysics, Karl-Schwarzschild-Str. 1, 85748 Garching, Germany \and 
Ludwig-Maximilians-Universit\"{a}t M\"{u}nchen, Geschwister-Scholl-Platz 1, 80539, M\"{u}nchen, Germany \and 
CSIRO Astronomy and Space Science, 26 Dick Perry Avenue, Kensington, WA 6151, Australia \and 
Dept. of Earth and Space Sciences, Chalmers University of Technology, Onsala Space Observatory,  439 92, Onsala, Sweden \and 
Ru{\dj}er Bo\v{s}kovi\'{c} Institute, Bijeni\v{c}ka cesta 54, 10000 Zagreb, Croatia \and 
Jodrell Bank Centre for Astrophysics, Alan Turing Building, School of Physics and Astronomy, The University of Manchester, Oxford Road, Manchester, M13 9PL, UK \and 
Leiden Observatory, Leiden University, PO Box 9513, 2300 RA Leiden, The Netherlands \and 
International Centre for Radio Astronomy Research - Curtin University, GPO Box U1987, Perth, WA 6845, Australia 
}

\date{Received 13 September 2016 / Accepted 30 November 2016} 

\abstract{Magnetic fields pervade the interstellar medium (ISM), but are difficult to detect and characterize. The new generation of low-frequency radio telescopes, such as the Low Frequency Array (LOFAR: a Square Kilometre Array-low pathfinder), provides advancements in our capability of probing Galactic magnetism through low-frequency polarimetry. Maps of diffuse polarized radio emission and the associated Faraday rotation can be used to
infer properties of, and trace structure in, the magnetic fields in the ISM. 
However, to date very little of the sky has been probed at high angular and Faraday depth resolution.

We observed a 5$^{\circ}$ by 5$^{\circ}$ region centred on the nearby galaxy IC342 ($\ell = 138.2^{\circ}, \, b = +10.6^{\circ}$) using the LOFAR High Band Antennas in the frequency range 115--178 MHz. We imaged this region at $4\farcm5 \times 3\farcm8$ resolution and performed Faraday tomography to detect foreground Galactic polarized synchrotron emission separated by Faraday depth (different amounts of Faraday rotation). Our Faraday depth cube* shows rich polarized structure, with up to 30 K of polarized emission at 150 MHz. We clearly detect two polarized features that extend over most of the field but are clearly separated in Faraday depth. 

Simulations of the behaviour of the depolarization of Faraday-thick structures at such low frequencies show that such structures would be too strongly depolarized to explain the observations. These structures are therefore rejected as the source of the observed polarized features. Only Faraday thin structures will not be strongly depolarized at low frequencies; producing such structures requires localized variations in the ratio of synchrotron emissivity to Faraday depth per unit distance. Such variations can arise from several physical phenomena, such as a transition between regions of ionized and (mostly) neutral gas. 

We conclude that the observed polarized emission is Faraday thin, and propose that the emission originates from two mostly neutral clouds in the local ISM. Using maps of the local ISM to estimate distances to these clouds, we have modelled the Faraday rotation for this line of sight and estimated that the strength of the line of sight component of magnetic field of the local ISM for this direction varies between $-0.86$ and +0.12 $\upmu$G  (where positive is towards the Earth). We propose that this may be a useful method for mapping magnetic fields within the local ISM in all directions towards nearby neutral clouds.}


\begin{document}
\maketitle

\section{Introduction}\label{sec:intro}

The ISM contains gas in a variety of physical conditions (cold molecular, cold and warm neutral atomic, warm and hot ionized), a population of relativistic particles (cosmic rays), dust, and an ambient magnetic field. Many aspects of the ISM are difficult to study because most of the tracers for the various components are difficult to measure, often require ancillary data, and often give integrated or average values for the physical parameters being estimated. The detection and estimation of magnetic fields in the ISM introduces the additional complication that the observational tracers also depend on one of the matter components.

Synchrotron polarization and Faraday rotation are often measured together to provide complementary information on interstellar magnetic fields. Synchrotron emission (and its polarization) traces the component of the magnetic field perpendicular to the line of sight but also depends on the cosmic ray properties. Faraday rotation provides information on the parallel component of the magnetic field along the line of sight but also depends on the thermal electron density.

Diffuse synchrotron polarization at low frequencies has shown a great deal of structure that has no counterpart in total intensity \citep[e.g.,][]{Wieringa93,Gray98,Haverkorn04}. This structure can be introduced both by fluctuations in the polarization at the emitting source, and by variations in the amount of Faraday rotation along the line of sight. As a result, these structures can provide unique information on the magnetic fields in the ISM.

The amount of information that can be extracted from polarization observations has been greatly increased by the development of the rotation measure (RM) synthesis technique \citep{Burn66, Brentjens05}, which can separate polarized emission by the degree of Faraday rotation it has experienced. The amount of Faraday rotation (i.e., the extent to which the polarization position angle has rotated between the emission source and the receiver) is the product of the observing wavelength squared ($\lambda^2$) and the Faraday depth ($\phi$) which is defined as 
\begin{equation}
\phi(d) = 0.812\; {\rm rad \, m^{-2}} \int_{d}^{0} \left( \frac{n_\mathrm{e}}{{\rm cm^{-3}}}\right) \left( \frac{\vec{B}}{{\rm \upmu G}} \right) \cdot \left(\frac{\vec{dl}}{{\rm pc}} \right),
\end{equation}
where $n_\mathrm{e}$ is the number density of free electrons, $\vec{B}$ is the magnetic field, $\vec{dl}$ is a differential element of the radiation path, and the integral is taken over the line of sight from a distance $d$ to the receiver. Polarized emission detected at different Faraday depths can be used to reconstruct the magnetic field  along the line of sight. This technique can be applied to a region of the sky to produce data cubes showing the distribution of diffuse polarized emission in position on the sky and in Faraday depth. We refer to the production and analysis of these data as {\it Faraday tomography}.
The resolution in Faraday depth depends on the range of $\lambda^2$ covered by the observations, so observations at low frequencies and with high fractional bandwidth give better resolution.

Faraday tomography of the Milky Way has been done previously with several datasets from the Westerbork Synthesis Radio Telescope (WSRT) \citep[e.g.,][]{Brentjens11, MarcoFan}, LOFAR \citep{Jelic14,Jelic15}, and the Murchison Widefield Array \citep{Lenc16}, as well as at higher frequencies with the 26m telescope at the Dominion Radio Astrophysical Observatory \citep{Wolleben10}. Many of these studies have been focused on characterizing features which appear in polarized emission and have no apparent counterpart in total intensity (which can occur when the total intensity emission is spatially smooth and is filtered out by an interferometer). Some studies have proposed models for the Faraday rotation of the diffuse polarized emission \cite[e.g., the screen and bubble model of ][]{MarcoFan}, identifying regions of emission and Faraday rotation along the line of sight and estimating the magnetic field strengths, electron densities, and distances associated with the Faraday rotation.

In this paper, we report on LOFAR observations of Galactic diffuse polarized emission towards the nearby galaxy IC342 and the results of performing Faraday tomography on these observations. In Sect. \ref{sec:data_processing} we describe the observations and their processing. We present the resulting Faraday depth cubes and describe the features observed in Sect. \ref{sec:cubes}. We follow this with a model of the magnetic field along these lines of sight in Sect. \ref{sec:model} and discuss the interpretation and limitations of this model in Sect. \ref{sec:discussion}. Our conclusions are summarized in Sect. \ref{sec:conclusions}.

\section{Observations and data processing} \label{sec:data_processing}
Our data consist of two observations with the LOFAR high-band antennas \citep[HBA, for full details on LOFAR's design see][]{vanHaarlem2013}. The first observation was taken from 2013-02-02/15:50 to 20:53 UTC, while the second was taken from 2013-03-13/22:21 to 2013-03-14/03:56 UTC. The full LOFAR `Dutch array', consisting of 48 core and 13 remote stations, was used in the HBA\_DUAL\_INNER mode. Each observation consisted of 19 pairs of pointings, with each pair containing a 120-second observation of the flux calibrator, 3C147 ($\alpha = 05^{\mathrm{h}}42^{\mathrm{m}}36^{\mathrm{s}}1, \, \delta = +49^{\circ}51'07''$), followed by a 720-second observation of the target field, centred on galaxy \object{IC342} ($\alpha = 03^{\mathrm{h}}46^{\mathrm{m}}48^{\mathrm{s}}5, \, \delta = +68^{\circ}05'46''$; $\ell = 138.1726^{\circ}, \, b = +10.5799^{\circ}$). The observed bandwidth was divided into 324 subbands, each with a bandwidth of 0.1953 MHz further divided into 64 channels, providing contiguous frequency coverage from 114.952 MHz to 178.233 MHz. An integration time of 1 second was used for all pointings, resulting in a raw data volume of about 40 TB.

We performed radio frequency interference (RFI) detection and flagging using the AOflagger algorithm \citep{Offringa2012}, which was applied to the data in three passes: on the raw data, after initial averaging, and after amplitude calibration.
Before the initial RFI flagging, we flagged the two lowest and two highest channels in each subband, as these channels are generally affected by the bandpass edges of the polyphase filter. After the initial RFI detection and flagging, we averaged the data in time and frequency to 6 seconds and 8 channels per subband (24.413 kHz bandwidth per channel), to reduce the data volume to approximately 1 TB.
The possibility of contamination by the bright `A-team' sources (Cas A, Cyg A, Vir A, Her A, and Tau A) was checked by simulating the contribution to the visibilities using the Blackboard Selfcal System \citep[BBS, ][]{Pandey2009} and found to be minor except for a few baselines at particular times. The `demixing' algorithm of \citet{vanderTol2007} was not used, and those baselines and time intervals that showed significant A-team signal were flagged. Before calibration, the stations CS013HBA0 and CS013HBA1 were completely flagged as the antennas in these stations were rotated with respect to the rest of the array.

The calibration target, 3C147, was calibrated with the flux model from \citet{Scaife2012}, using the BBS software, independently for each subband and 2-minute calibration pointing. The resulting gain amplitude solutions were interpolated in time and applied to the target field. For the phase calibration of the target field, a sky model was made using the LOFAR global sky model (GSM), which was made by combining the catalogs from the NRAO VLA Sky Survey \citep[NVSS]{Condon1998}, the Westerbork Northern Sky Survey \citep[WENSS]{Rengelink1997}, and the VLA Low-Frequency Sky Survey redux \citep[VLSSr]{Lane2014}. Phase calibration was performed on groups of 9 subbands, to improve the signal-to-noise ratio of the solutions.

No self-calibration was applied to the data. We found that the direction-independent phase calibration produced good results for the shorter baselines across most of the target field; we achieved the best images by removing the remote stations more distant from the LOFAR core. We chose to phase calibrate and image using only the core stations and the nearest three remote stations (RS305, RS503, and RS205).

To accurately determine the Faraday depths, we removed the contribution of the ionosphere to the Faraday rotation, using the RMextract software\footnote{https://github.com/maaijke/RMextract/} written by Maaijke Mevius. This software calculates the ionospheric contribution by using the World Magnetic field Model (WMM)\footnote{https://www.ngdc.noaa.gov/geomag/WMM/DoDWMM.shtml}, maps of the total free electron content of the ionosphere from the Center for Orbital Determination in Europe (CODE)\footnote{http://aiuws.unibe.ch/ionosphere/}, and a model for the ionosphere to predict the Faraday rotation of the ionosphere for a given LOFAR observation. The observations were derotated by the predicted amount using the BBS software. The estimated systematic uncertainty in the Faraday depth correction is approximately 0.1--0.3 \radu \ \citep{Sotomayor13}.

Before imaging, the baselines between each pair of HBA sub-stations (e.g., CS002HBA0 and CS002HBA1) were removed, as these were observed to have significant instrumental cross-talk. Imaging was performed with the AWimager \citep{Tasse2013}. Images were made in Stokes $Q$ and $U$ for each channel, using robust weighting of 1.0 and including only baselines between 10 and 800$\lambda$. Station beam correction was applied within AWimager, and due to very low signal-to-noise in each image no cleaning was done.
This produced a frequency independent resolution of $4\farcm5 \times 3\farcm8$. This was done for all 2592 channels in the data set. After imaging, 110 channels were identified by manual inspection as being badly affected by noise or instrumental effects and were removed. The standard deviation of flux density at the center of each image was about 12 mJy PSF$^{-1}$ and almost independent of frequency, giving a theoretical band-averaged rms noise level of 0.24 mJy PSF$^{-1}$ at the center of the field. However, these values contain contributions from both the per-channel noise and the signal present in each channel, and so represent an over-estimate of the true noise in the data.

\begin{figure}[htbp]
\resizebox{\hsize}{!}{\includegraphics[width=\linewidth]{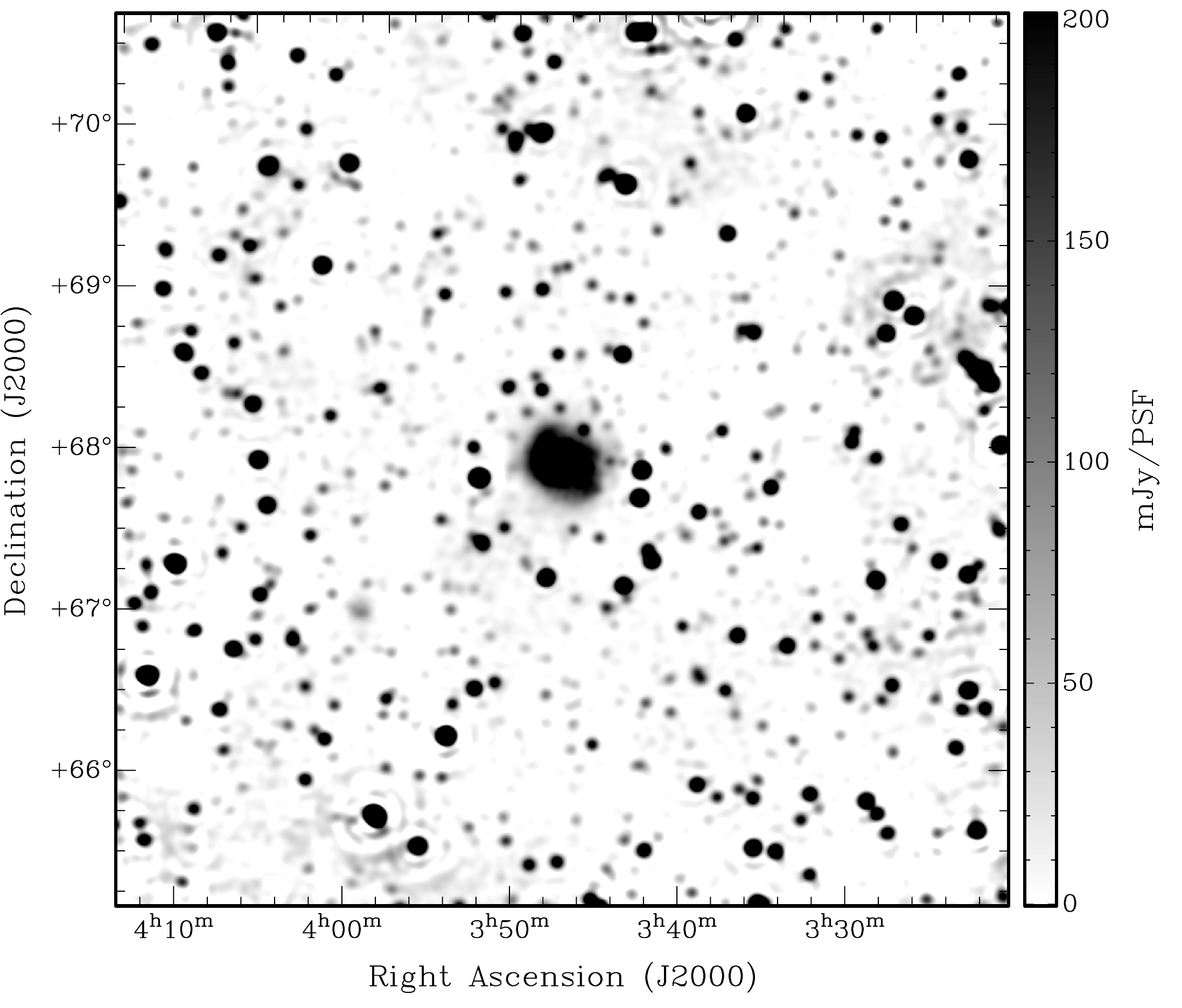}}
\caption{Total intensity map of the IC342 field. The nearby spiral galaxy IC342 appears prominently in the center, the giant double radio galaxy \object{WNB 0313+683} appears on the right, and the dwarf galaxy \object{UGCA 86} ($\alpha = 03^{\mathrm{h}}59^{\mathrm{m}}49^{\mathrm{s}}4, \delta = +67^{\circ}08'38''$) appears faintly below and left of center. In Galactic coordinates, the center of the field is at $\ell = 138.2^{\circ}, \, b = +10.6^{\circ}$. The resolution is $4\farcm5 \times 3\farcm8$.}
\label{fig:StokesI}
\end{figure}

In addition, a Stokes $I$ image was produced to search for polarized point sources. This image was produced using the full bandwidth and time range of the observations, and used the same baseline selection as the polarization images. The resulting image is shown in Fig. \ref{fig:StokesI}. A detailed analysis of the Stokes $I$ emission from IC342 is deferred to a future paper.

The pyRMsynth\footnote{https://github.com/mrbell/pyrmsynth} software package was used to perform the Faraday tomography. The frequency coverage of the data produced a Faraday depth resolution of 0.9 \radu, a maximum scale of 1.1 \radu, and sensitivity to Faraday depths in the range $|\phi| < 2200$ \radu, as calculated from equations 61-63 of \citet{Brentjens05}. However, these equations are only applicable with the criterion that $|\phi|\Delta\lambda^2 \ll 1$, which is only satisfied for $|\phi| \ll 350$ \radu \ at the lowest frequency. Since our field is outside of the Galactic plane, we do not expect emission at large Faraday depths, so this criterion should not be violated. The small difference between the resolution and the maximum scale means that we are not able to resolve Faraday depth structure; features broader than the maximum scale will be strongly depolarized and thus filtered out, while features narrower than the resolution will appear as unresolved peaks. The consequences of this are discussed in Sect. \ref{sec:theory} and Appendix \ref{app:slabs}.

Channel weights were applied inside pyRMsynth, and were made equal to the inverse square of the rms noise in each image (analogous to natural weighting in radio interferometry). Uniform channel weighting was also tested and found to produce insignificant differences in the final Faraday cubes. The restoring beam used in RM-cleaning was a Gaussian fitted to the rotation measure spread function (RMSF, the response function introduced by limited sampling in the wavelength domain), with a fitted standard deviation of 0.37 \radu \ (corresponding to a FWHM of 0.87 \radu, in agreement with the theoretical resolution above). RM-CLEAN \citep{Heald09} was applied to each cube, down to a threshold of 2 mJy PSF$^{-1}$ RMSF$^{-1}$. No correction for the spectral index of the emission was applied, as the diffuse flux was not detected in total intensity to determine the appropriate spectral index; this may introduce a small error in the polarized intensities of order 2--5\% \citep{Brentjens05}. Cubes were also made without applying RM-CLEAN, and found to have no significant differences to those with RM-CLEAN.

During calibration, polarization leakage from Stokes $I$ into Stokes $Q$ and $U$ was not corrected for, as at the time of processing no method had been developed to determine this correction and the effects on the data were judged minor enough to not merit reprocessing. The leakage produces apparent polarization at the location of all Stokes $I$ sources. The leakage is frequency-independent, so the spurious polarization appears at 0 \radu \ in Faraday depth. However, the ionospheric Faraday rotation correction causes all the polarization to be shifted in Faraday depth by the opposite of the predicted ionospheric Faraday depth, to remove the ionospheric contribution. By doing so, the astrophysical signal was moved to the correct Faraday depth, and the instrumental polarization was moved away from 0 \radu. Since the ionospheric correction was time-variable, the leakage is `corrected' to different values for each time. In the resulting Faraday cubes the leakage is then smeared out over a range of Faraday depths corresponding to the (negative of the) range of values of the ionospheric correction. For these data, this correction ranged from 0.2 to 1.1 \radu, so the instrumental polarization was shifted to between $-1.1$ and $-0.2$ \radu \ and, due to convolution with the RMSF, appears in the cube slices between approximately $-1.5$ and +0.5 \radu.

Two Faraday depth cubes were produced: a finely-sampled cube, covering Faraday depths from $-25$ to 25 \radu \ in steps of 0.25 \radu, and a more coarsely-sampled cube from $-100$ to 100 \radu \ in steps of 0.5 \radu. The catalog of \citet{Taylor09} contains polarized sources with rotation measures between $-$70 \radu \ and $+$23 \radu \ in this region of the sky, so we did not expect any diffuse emission or point sources with Faraday depths beyond $\pm 100$ \radu. The final noise in the cubes was determined on a pixel-by-pixel basis by masking out Faraday depths between $-20$ and +20 \radu \ (where most of the signal was expected), constructing a histogram of the polarized intensity distribution for the remaining (empty) Faraday spectrum, and fitting a Rayleigh distribution with a least-squares solver. The Rayleigh distribution represents the distribution of polarized intensity when the distributions of Stokes $Q$ and $U$ are both Gaussian. We found that this method gave similar values to fitting Gaussians to the noise distributions in $Q$ and $U$, with the advantage of using all the data in a single fit. The resulting noise (expressed as Rayleigh $\sigma$ parameter, which is equivalent to the Gaussian $\sigma_{\mathrm QU}$ of the Stokes $Q$ and $U$ distributions) was position dependent (due to beam correction), and ranged from 0.2 mJy PSF$^{-1}$ RMSF$^{-1}$ near the center of the field (in agreement with the band-averaged noise estimate above) to approximately 2 mJy PSF$^{-1}$ RMSF$^{-1}$ in the lower left and upper right corners (3.5 degrees from the phase center). At the band center, 146.6 MHz, the conversion from flux density to brightness temperature is 0.924 K (mJy PSF\inv)\inv, from equations 9-25 and 9-26 of \citet{Wrobel99}.

\section{Faraday depth cubes}  \label{sec:cubes}
\begin{figure*}[htbp]
\includegraphics[width=16cm]{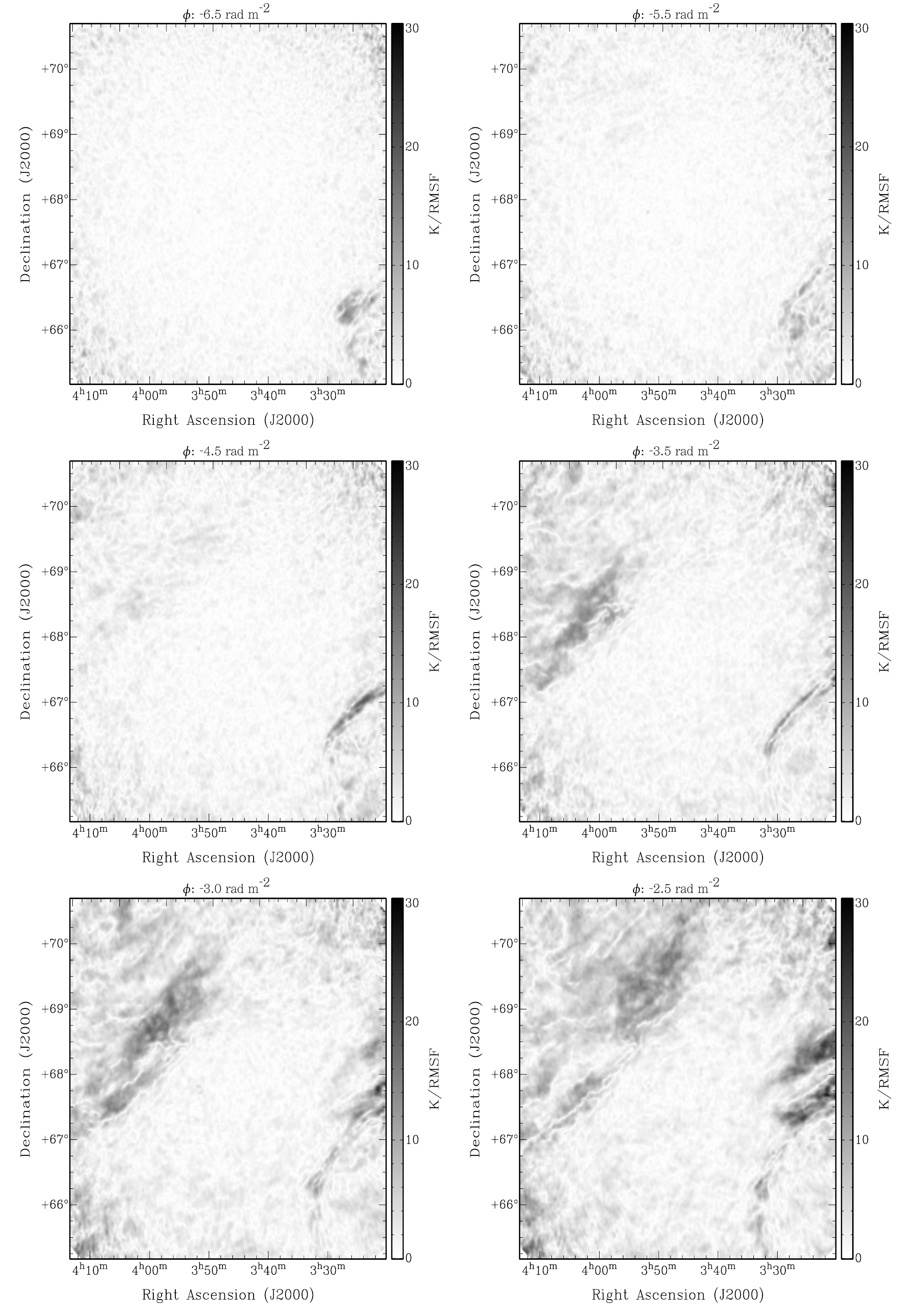}
\caption{Selected slices from the finely-sampled Faraday depth cube, showing the polarized intensity at different Faraday depths from $-$6.5 \radu \ to $-$2.5 \radu. A bright polarized diffuse feature can be seen entering the field from the top left and bottom right corners. The resolution is $4\farcm5 \times 3\farcm8$.}
\label{fig:slices-1}
\end{figure*}

\begin{figure*}[htbp]
\includegraphics[width=16cm]{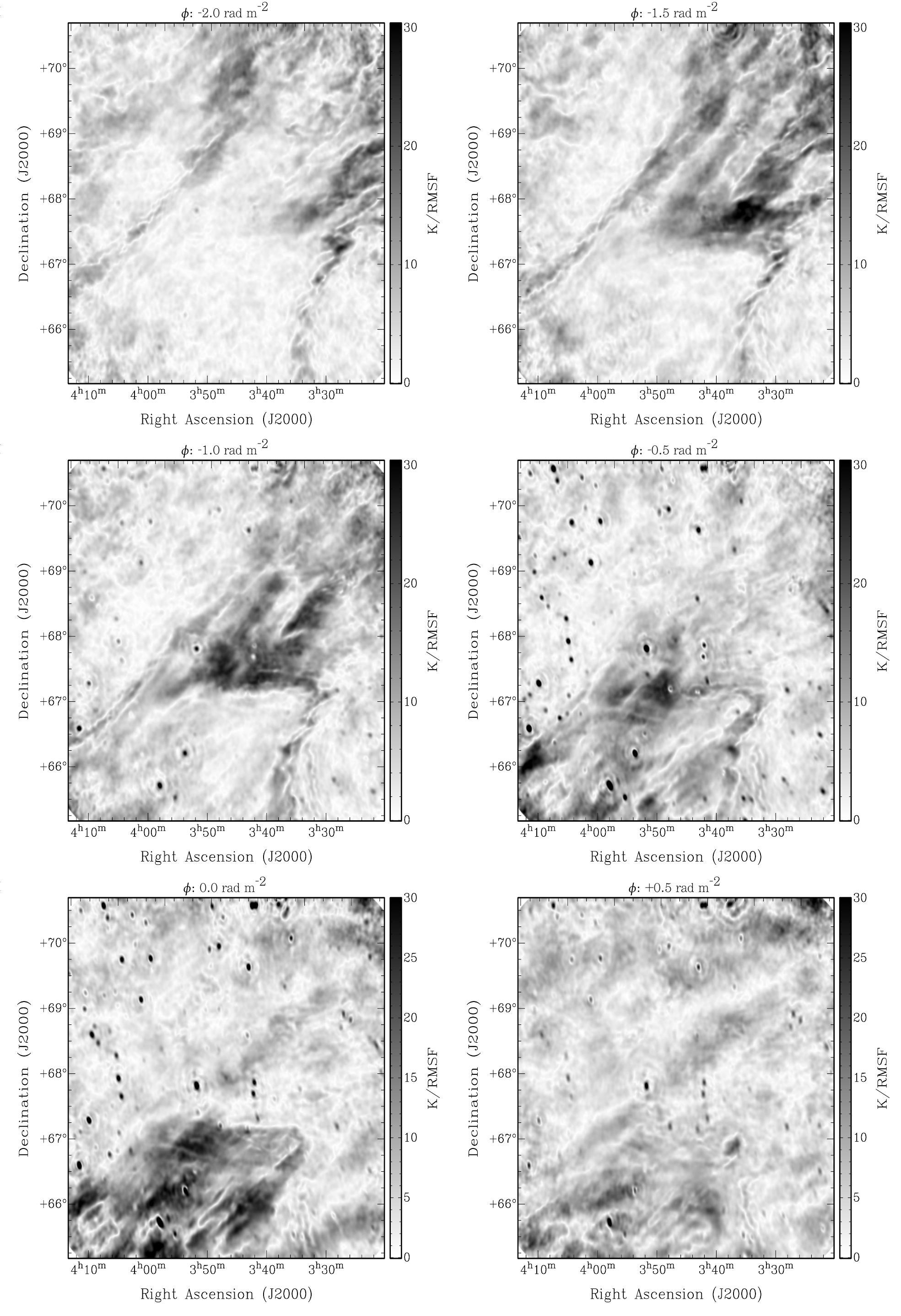}
\caption{As Fig. \ref{fig:slices-1}, more slices from the same cube, from $-$2 \radu \ to +0.5 \radu. The bright polarized feature can be seen to move through the center of the frame and towards the lower left corner. The polarization leakage from Stokes $I$ into $Q$ and $U$ can be seen at Faraday depths between $-1$ and +0.5 \radu.}
\label{fig:slices-2}
\end{figure*}

\begin{figure*}[htbp]
\includegraphics[width=16cm]{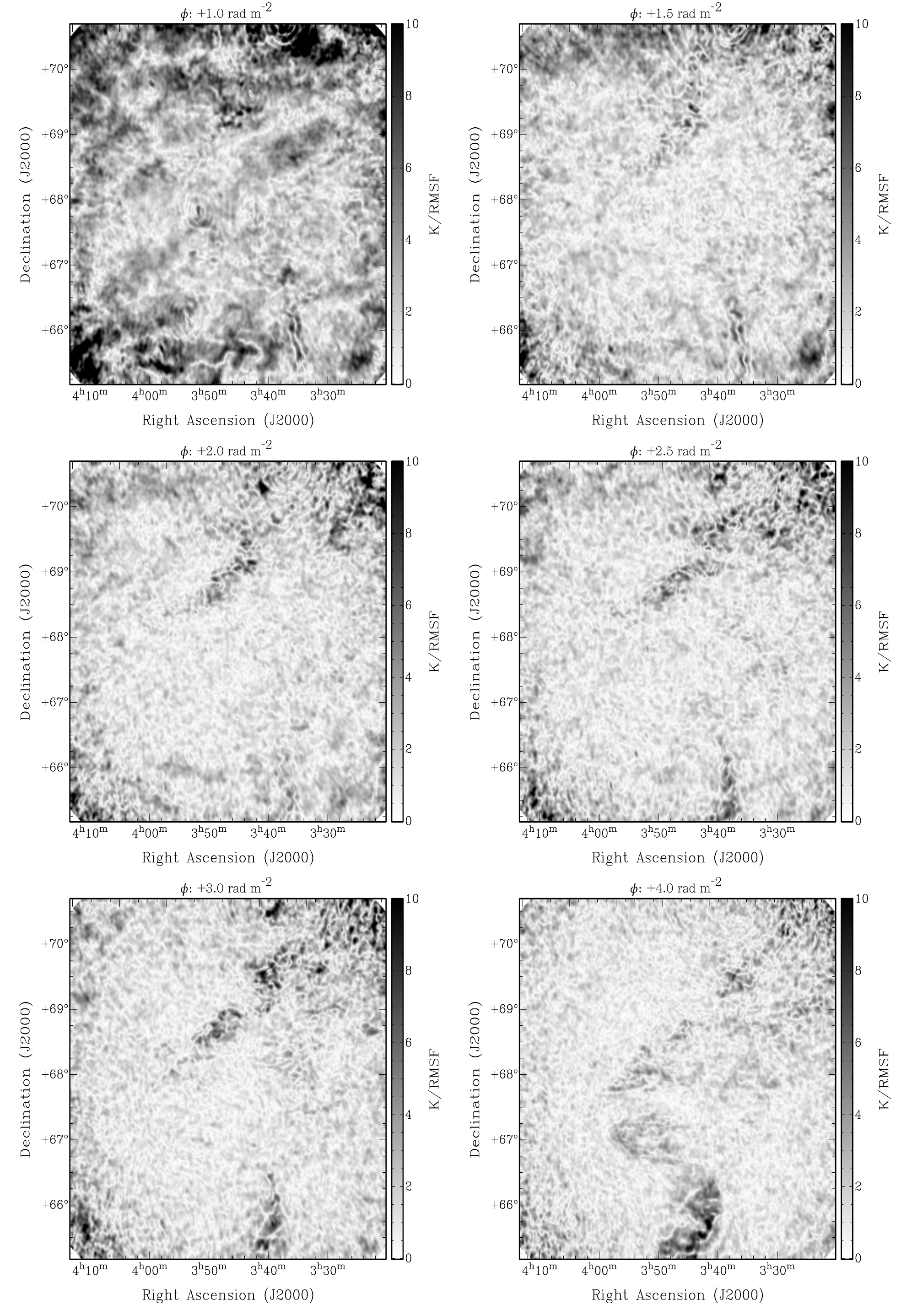}
\caption{As Fig. \ref{fig:slices-1}, more slices from the same cube, from +1 \radu \ to +4.0 \radu. The intensity scale has been adjusted to show the faint emission more clearly. The bright polarized feature fades away, and a second, fainter feature emerges in the top and bottom of the field, moving towards the lower left.}
\label{fig:slices-3}
\end{figure*}

\begin{figure*}[htbp]
\includegraphics[width=16cm]{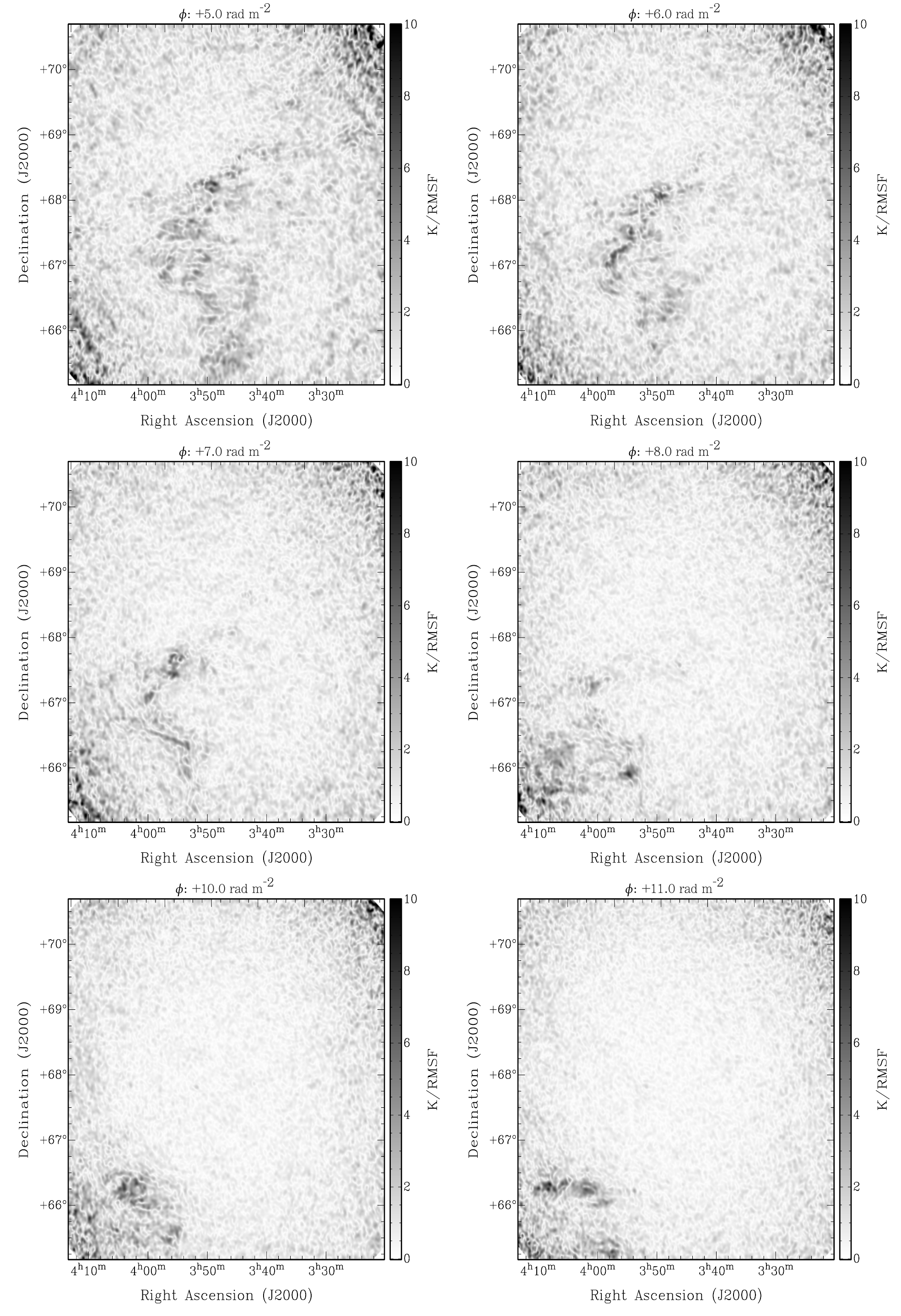}
\caption{As Fig. \ref{fig:slices-1}, more slices from the same cube, from +5 \radu \ to +11 \radu. The intensity scale has been adjusted to show the faint emission more clearly. The faint polarized feature moves through the center towards the lower left.}
\label{fig:slices-4}
\end{figure*}

\begin{figure*}[htbp]
\includegraphics[width=17cm]{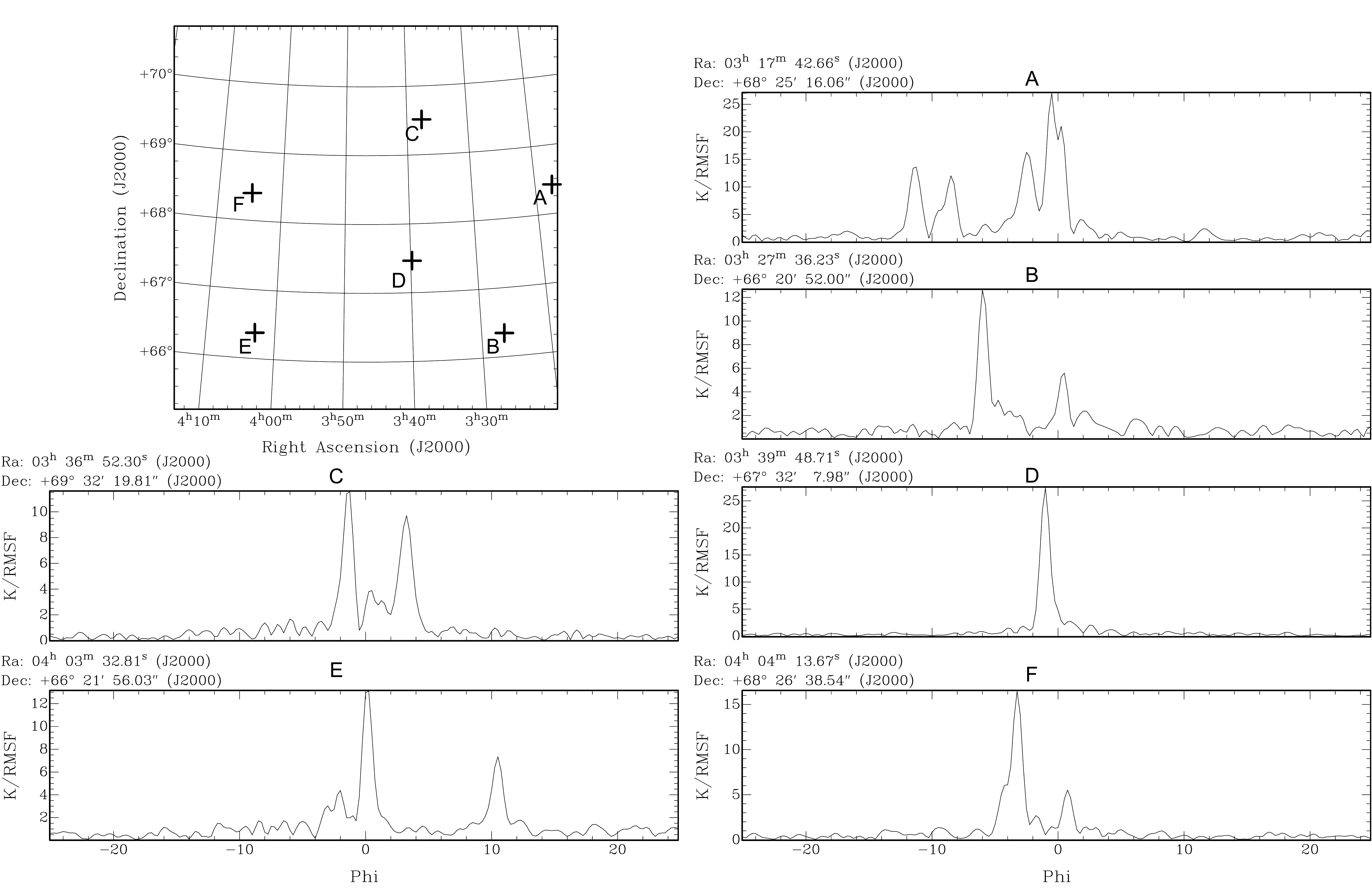}
\caption{Faraday depth spectra for selected lines of sight. The top left panel shows the locations of the lines of sight. Panel A is at the location of the polarized emission from the giant radio galaxy WNB 0313+683, which shows two Faraday depth components at $-11.4$ and $-8.6$ \radu \ (which overlap in angular position at this resolution), a Stokes $I$ leakage feature at $-0.5$ \radu, and a diffuse emisison peak at $-3$ \radu. Other panels are at locations containing only diffuse emission, and show either one (D) or two (B,C,E,F) clear peaks.}
\label{fig:spectra_1}
\end{figure*}

In this section we present and describe the resulting Faraday depth cubes. Figs. \ref{fig:slices-1} through \ref{fig:slices-4} show images of polarized intensity extracted from the Faraday cubes, which were selected to show the interesting features in the cube. Fig. \ref{fig:spectra_1} shows some sample Faraday depth spectra for different positions in the cube.

In broad terms, the observed polarized emission can be divided into four components:
\begin{itemize}
\item instrumental polarization leakage, appearing between $-1.5$ \radu \ and +0.5 \radu, as discussed in Sect. \ref{sec:data_processing};
\item unresolved polarized sources, most likely background radio galaxies, observed between $-30$ and $-8$ \radu;
\item a diffuse emission feature, with a complex morphology that covers most of the field, between $-$7 \radu \ and +3 \radu, with a typical polarized brightness of 30 K;
\item a second, fainter, diffuse emission feature, with a different morphology, between +1.5 \radu \ and +11 \radu, with a typical polarized brightness of 10 K.
\end{itemize}

\subsection{Polarized background sources}

Three unresolved polarized background sources were detected in the 5$^{\circ}$ by 5$^{\circ}$ field. Two of these coincide with locations in the double radio galaxy WNB 0313+683, with different Faraday depth values and slightly different positions. The third is a single radio source, NVSS J041445+690108. All three were matched with sources in the \citet{Taylor09} RM catalog. Table \ref{table:pointsources} gives the measured parameters for these sources.

\begin{table*}
\caption{Measured parameters of the polarized background sources}
\label{table:pointsources} 
\centering
\begin{tabular}{c c c c c c }
\hline\hline
$\alpha_{J2000}\tablefootmark{a}$ & $\delta_{J2000}$\tablefootmark{a} & I(\@150 MHz)\tablefootmark{b} & PI(\@150 MHz)\tablefootmark{c} & RM\tablefootmark{d} & NVSS RM \tablefootmark{e} \\ 
$\phantom{.}$ [h m s] & [d m s] & [mJy PSF\inv] & [mJy PSF\inv] & [\radu] & [\radu]\\
\hline 
04 14 45$\pm$2 & 69 01 14$\pm$9 & \phantom{0}562$\pm$\phantom{0}5 & \phantom{0}8.0$\pm$0.5 & $-28.$6 $\pm$ 0.05 & $-32.9$$\pm$$15.8$ \\
03 17 47$\pm$1 & 68 24 54$\pm$6\tablefootmark{f} & 1740$\pm$10 & 15.4$\pm$0.7 & $-11.4 \pm$ 0.05 & $-12.9$$\pm$$\phantom{0}4.9$ \\
03 17 40$\pm$1 & 68 24 03$\pm$6\tablefootmark{f} & 1320$\pm$10 & 14.6$\pm$0.7 & \phantom{0}$-8.6 \pm$ 0.05 & $-12.8$$\pm$$\phantom{0}4.7$ \\
\hline
\end{tabular}
\tablefoot{
\tablefoottext{a}{Position from fitting the source in polarized intensity.}
\tablefoottext{b}{Observed intensity at the pixel closest to the fitted position.}
\tablefoottext{c}{Polarized intensity, found by fitting a 3D Gaussian to the source.}
\tablefoottext{d}{Rotation measure, found by fitting a 3D Gaussian to the source. The ionospheric Faraday rotation correction introduces an additional systematic error of about 0.1--0.3 \radu.}
\tablefoottext{e}{Rotation measure from the catalog of \citet{Taylor09}.}
\tablefoottext{f}{These sources are at the position of WNB 0313+683.}
}
\end{table*}

A consequence of the instrumental polarization is that any polarized sources with Faraday depths between $-1.5$ and +0.5 \radu \ cannot be separated from the leakage, and currently cannot be identified. Of the 45 sources in this field with cataloged RMs from Taylor et al. 2009, there is only one source with an RM value within this range (11 additional sources are within 1$\sigma$ of this range). Since we detected only 3 polarized sources in the accessible Faraday depth range of our data, out of the 44 known polarized sources in this range, we conclude that it is unlikely that another polarized source is hidden inside the instrumental leakage signal.

Due to the small number of sources, we defer a detailed analysis of these polarized sources to a planned follow-up paper, which will use this and other LOFAR observations to construct a much larger and statistically useful sample of low-frequency polarized sources.

No obvious polarization was observed at the location of IC342, other than the instrumental polarization leakage from Stokes $I$. A careful upper limit on the polarization of IC342 at this frequency is deferred to a future paper where the data will be reprocessed at full resolution, to reduce the possible effects of beam depolarization.

\subsection{Diffuse polarized emission}

We divide the diffuse polarized emission into two features, based on the morphology and range of Faraday depths. Both have similar large-scale structure in Faraday depth, but are displaced from each other by several \radu. The first feature covers a Faraday depth range between about $-7$ and +3 \radu \ (Figs. \ref{fig:slices-1} to \ref{fig:slices-3}), and consists of diffuse emission across the entire field. The lowest Faraday depths occur at the lower right and upper left corners, with a gradient towards the center and upper right. Around Faraday depth $-2$ to $-1$ \radu \ the center and upper right become filled with emission, and two filamentary `arms' extend to either side of the lower left corner. From $-1$ to +1 \radu \ there is a strong gradient, with the emission sharply transitioning from the upper right to the lower left. At Faraday depths greater than +1 \radu, there is some remaining diffuse emission in the lower left corner, which remains present to at least +3 \radu, but it is difficult to determine where exactly the emission ends as the edges of the cube are significantly affected by noise (due to the beam correction). The morphology of this emission matches up very well with the observations of \citet{MarcoFan}, which overlap the lower right corner of our field.

This emission feature also contains a number of long, nearly straight depolarization canals. These canals appear to have a preferred axis (towards the lower-left and upper-right corners), which appears to be aligned well with the Galactic plane (lines of constant Galactic latitude also run from the lower-left to the upper-right). Since we did not use CLEAN on the individual $Q$ and $U$ channel images, these canals are not artifacts of the type described by \citet{Pratley16}, but reflect real structure in the emission (albeit affected by the resolution of the observations). Further investigation into the significance and possible interpretations of this are left for a follow-up analysis.

The second, fainter, diffuse feature covers a significant fraction of the field at higher Faraday depths, from +1.5 to +11 \radu \ (Fig. \ref{fig:slices-3} and \ref{fig:slices-4}). This feature has a similar trend to the first: at the lowest Faraday depths it occurs in the top left and lower right corners, with a gradient towards the center and upper right with increasing Faraday depth values. At Faraday depths between +3.5 and +5 \radu \ it can be seen to fill much of the center and upper right of the frame, and transitions sharply towards the lower left between +5 and +11 \radu. This feature shows very similar behaviour in the Faraday depth gradients to the first, but the structure of the emission (i.e. extent of emission, and locations of bright regions and canals) is different between the two. This suggests that the gradient is the result of a large-scale foreground Faraday-rotating screen in front of both emission features, while the structure in the polarized intensity is unique to each source of diffuse polarized emission.

One concern when interpreting Faraday spectra is the risk of mis-identifying instrumental artifacts as real features. This can occur, for example, when the RMSF sidelobes of two emission features interact to produce a third, artificial feature. We conclude that this is not the case for the weaker emission feature we see here, and also that the fainter feature is not a sidelobe of the brighter feature, for three reasons. Firstly, through most of the field, there is no second bright feature that would mix with the brighter diffuse emission feature. It is possible that the instrumental leakage and the real emission could mix and produce an apparent feature in the spectrum, but this would be more likely to occur at Faraday depths between the real emission and the leakage, not at higher Faraday depths (an interaction like this between the two emission features could explain the small, 3 K RSMF\inv \ peaks seen around +1 \radu \ in spectrum C of Fig. \ref{fig:spectra_1}). Also, the leakage is mostly confined to the point sources, and would not be able to produce a spurious diffuse feature. Secondly, the first sidelobes in the RMSF are separated from the main lobe by $\pm$1.2 \radu, while the two diffuse emission features are observed to be separated by 4--10 \radu. At this separation, the RMSF sidelobes have a strength between 6\% and 4\% of the main peak, which is too small to explain the observed intensity of the second feature. Thirdly, the morphology between the two emission features shows significant differences, which can't be easily explained if the two features are related by some instrumental effect. For these reasons, we conclude that the fainter emission feature is real.

\begin{figure*}[htbp]  
{\includegraphics[width=17cm]{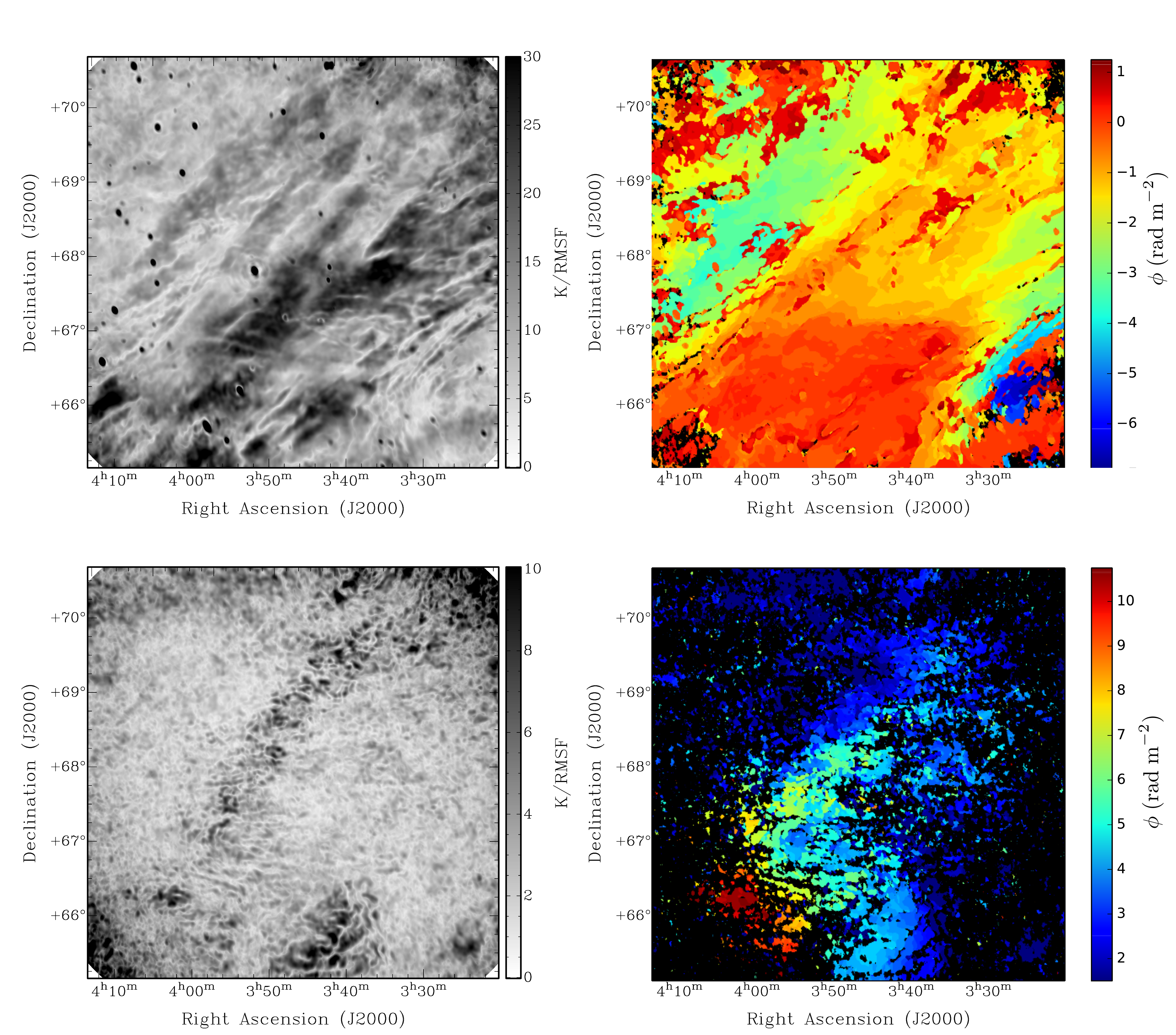}}
\caption{Left: Maps of the peak polarized intensity in selected Faraday depth ranges. Right: Maps of the Faraday depth of peak emission, for the same ranges in Faraday depth. Pixels with peak polarized intensity below 10 $\sigma_{\mathrm QU}$ are masked. Top panels: Faraday depths between $-7$ and +1.5 \radu. Bottom panels: Faraday depths between +1.5 and +11 \radu. The compact sources in the top panel are caused by the instrumental polarization.}
\label{fig:gradients}
\end{figure*}

From the morphology, each diffuse feature appears to be a single emission region distributed across a range of Faraday depths: each represents a connected sheet in the three-dimensional volume of the Faraday cube, smoothly varying in Faraday depth as a function of position on the sky. In Fig. \ref{fig:gradients} we show the Faraday depth and polarized intensity of each feature per pixel, by finding the peak polarized intensity in fixed Faraday depth ranges selected to pick out each feature. These maps demonstrate the same features observed in the individual slices: the two diffuse features have distinctly different morphologies in emission, but similar trends in Faraday rotation.

\section{Modeling the diffuse Galactic emission}   \label{sec:model}
In this section, we present a physical model that describes the main features of the diffuse emission described above. To do so, we first account for the effects of incomplete wavelength coverage on the Faraday spectrum, and then consider possible physical configurations that might produce the observations given these effects.

\subsection{Properties of low-frequency RM synthesis} \label{sec:theory}
Since RM synthesis is a Fourier transform-like process, the reconstruction of the Faraday spectrum is affected by filtering due to incomplete sampling of the $\lambda^2$ domain. By analogy to radio interferometry, the dirty beam is represented by the RMSF, which is convolved with the actual Faraday spectrum to give the measured spectrum. The effects this has on the observed spectrum, especially the resulting limits to the information in a Faraday depth spectrum, have been studied by several authors \citep[e.g.,][and references therein]{Brentjens05,Beck12}. One such effect, which becomes very constraining at low frequencies, is the loss of sensitivity to broad structures in the Faraday spectrum (which are often called `Faraday thick' features, although this term is often tied to the Faraday depth resolution of a given observation), directly analogous to how a lack of short baselines removes large-scale emission in interferometry. This can also be interpreted in terms of wavelength-dependent depolarization by considering the Fourier scaling property: making a feature broader in Faraday depth makes the transform of that function narrower in the $\lambda^2$ domain. Broader features in Faraday depth result in the polarization becoming more rapidly depolarized with increasing wavelength.\footnote{It is worth noting here that we are only discussing {\it depth depolarization} in an emitting and Faraday-rotating volume, and neglecting the effects of {\it beam depolarization} by a Faraday-rotating foreground, which has been studied by, e.g., \citet{Tribble91,Sokoloff98, Schnitzeler2015b}.}

\begin{figure}[!]
\resizebox{\hsize}{!}{\includegraphics{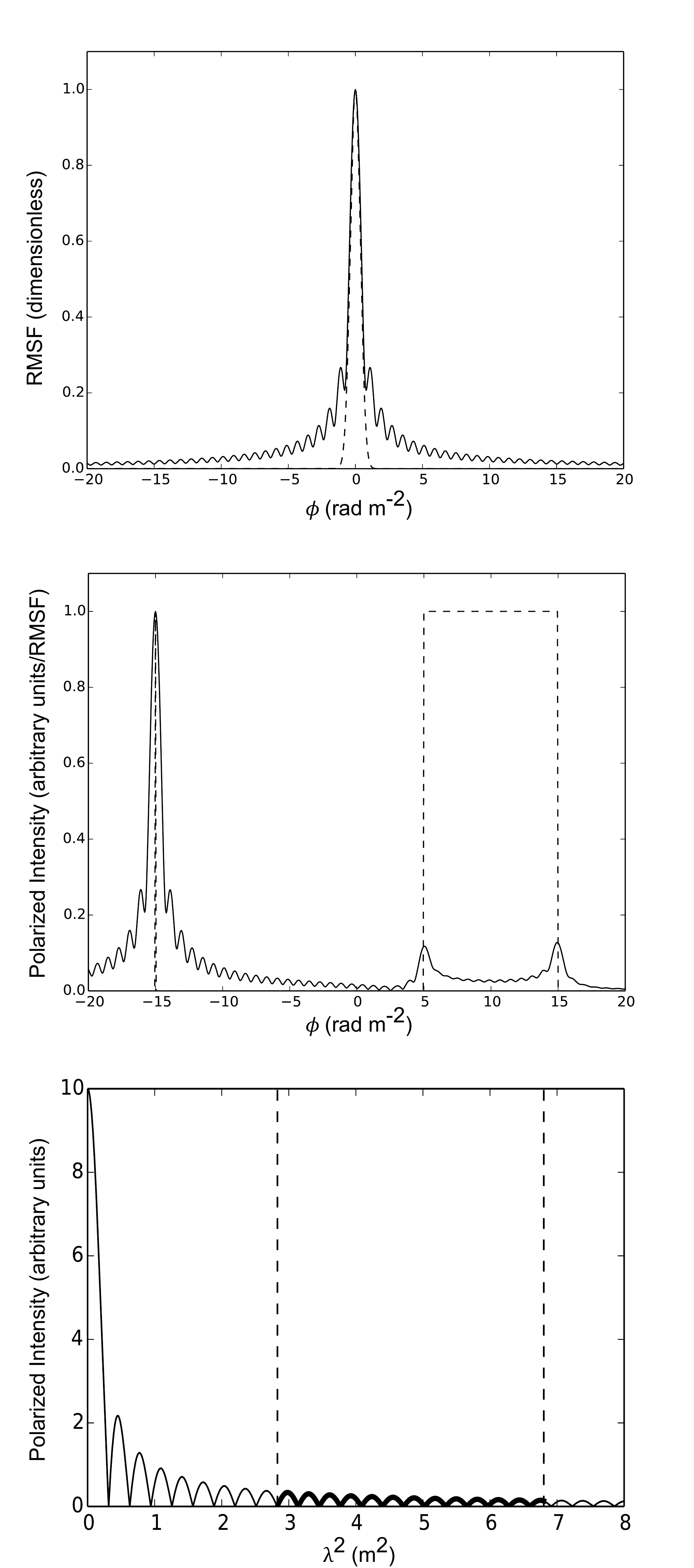}}
\caption{{\it Top:} Solid line: The RMSF for the frequency sampling of the IC342 observations. Dashed line: The Gaussian used as the restoring function in the RM CLEAN algorithm.
{\it Middle:} Dashed line: Input spectrum containing a delta function at $\phi = -15 \ $\radu \ and a Faraday slab between +5 and +15 \radu, both with amplitude of 1.
Solid line: Resulting spectrum using the frequency sampling of the IC342 observations. The Faraday slab is almost completely depolarized.
{\it Bottom:} The polarization as a function of $\lambda^2$ for the Faraday slab above. The wavelength range of the LOFAR HBA is between the two dashed lines and the simulated signal in this range is marked in bold.}
\label{fig:rmsimulator}
\end{figure}

This behaviour is demonstrated in Fig. \ref{fig:rmsimulator}, where we simulate a Faraday-thin component (modelled as a Dirac delta function) and a Faraday slab \citep[a top-hat or square pulse function in the Faraday spectrum, also called a Burn slab, ][]{Burn66}, using identical $\lambda^2$ coverage to our LOFAR observations. If the slab is significantly broader than the RMSF, the result is two peaks in the Faraday depth spectrum corresponding to the two edges of the tophat \citep{Brentjens05, Heald09, Beck12}. The measured amplitude of these two peaks, given our $\lambda^2$ coverage, is 12\% $\pm$ 1\% of the true amplitude  for all slabs thicker than about 2 \radu \ (see Appendix \ref{app:slabs} for a discussion of this value).

The result of the filtering in the observed spectrum is that smooth features are removed while narrow features or sharp edges (i.e., narrower than the width of the RMSF) are preserved in low-frequency observations. This has significant implications on the physical conditions that can be observed.
The key parameter that sets the amplitude in the Faraday spectrum, which we call $A_\phi$, is the ratio of polarized synchrotron emissivity to Faraday depth per unit distance, 
\begin{equation}\label{eq:amplitude}
A_\phi = \frac{p_0 \, \left( \frac{\varepsilon}{\mathrm{K \, pc^{-1}}} \right) }{0.812 \left( \frac{n_\mathrm{e}}{\mathrm{cm^{-3}}}\right) \left(\frac{| B_\parallel |}{\rm \upmu G} \right)}\, {\rm K\, (rad\, m^{-2})^{-1}},
\end{equation}
where the (total intensity) synchrotron emissivity, $\varepsilon$, depends on the cosmic ray electron density and the perpendicular magnetic field strength, and $p_0$ is the intrinsic polarization fraction of the emission. Sharp variations in this ratio, as a function of distance,  are one method to produce narrow or sharp features in the Faraday spectrum; it is the presence of these sharp variations that causes the Faraday slab to appear as two peaks (one peak where it sharply increases from zero to the slab's amplitude, and the second where it decreases back to zero). These variations can take the form of positive or negative changes to the Faraday spectrum amplitude; a sharp decrease in $A_\phi$ will produce a feature indistinguishable from a sharp increase after the broad components are filtered out.
Below, we consider some different physical processes that could produce such variations in $A_\phi$.

A localized enhancement in the perpendicular magnetic field, such as that produced by the shock of an expanding supernova remnant, will create a region of enhanced synchrotron emission. The limited depth of such a shock could very naturally produce a sharp feature in the Faraday spectrum, which may not depolarize much if the total Faraday depth produced inside the shock is less than the width of the RMSF. A diminishment in the perpendicular magnetic field would produce a similar (negative) feature in the Faraday spectrum.

The intrinsic polarization fraction, which is determined by how ordered the magnetic field is in the emitting region, may also vary and affect the Faraday spectrum amplitude. A region with a more ordered field or a more isotropic field will produce stronger or weaker polarization, respectively. A shock oriented perpendicular to the line of sight can make the magnetic field more ordered (parallel to the shock surface), giving the magnetic field in the region of the shock a preferred orientation and enhancing the polarization fraction.

The parallel component of the magnetic field could also be varied, either by an enhancement or diminishment. An enhancement would increase the strength of the Faraday rotation, which would decrease $A_\phi$, while a diminishment would have the opposite effect. A region where the parallel magnetic field component changes sign will produce a very sharp feature in the Faraday spectrum, called a Faraday caustic \citep{Bell11,Beck12}. Faraday caustics are strong candidates for detection at low frequencies, as they produce sharp, high-amplitude features that should not be strongly depolarized at low frequencies.

Finally, the free electron density can be varied, with the same effects as changes in the parallel magnetic field. Sharp localized changes in the free electron density can be associated with sharp density fluctuations, like shocks, and at interfaces between different gas phases of the ISM.

The different phases of the ISM have very different density and ionization conditions, leading to sharp changes in the electron density where a line of sight passes through regions containing different phases. Here we ignore the cold molecular phase and compact \ion{H}{II} regions, which occupy a very small fraction of the ISM and are not expected to be large enough to contain sufficient synchrotron-emitting volume to be detected in our data (also, these phases have not been observed in the lines of sight probed by our data). We confine our consideration to the 3 phases which occupy the bulk of the volume of the ISM: the warm ionized medium (WIM), warm neutral medium (WNM), and hot ionized medium (HIM).

The highest thermal electron densities are found in the WIM, which has been found to have electron densities of approximately 0.18-0.46 cm$^{-3}$ \citep{Ferriere01}.\footnote{Note that all the number densities we discuss are local, not volume averaged or multiplied by filling factors, as we want to consider the Faraday rotation occurring inside each phase.} The WNM has total number densities of approximately 0.1-0.6 cm$^{-3}$, and ionization fractions of a few percent \citep[typically inversely related to density; ][]{Wolfire95}. The resulting thermal electron densities in the WNM are approximately 0.01 cm$^{-3}$, although this can be significantly higher in the presence of additional ionization sources \citep{Jenkins13}. The HIM has a lower thermal electron density of approximately 0.005 cm$^{-3}$ \citep{Spangler09}.

Assuming a parallel magnetic field strength of 2 $\upmu$G and an electron density of 0.01 cm$^{-3}$, the WNM produces 0.016 \radu \ pc$^{-1}$ of Faraday rotation. For LOFAR observations like those presented here where the maximum scale is 1.1 \radu , this corresponds to a path length of 68 pc. From this, we predict that our observations should be fully sensitive to neutral regions of this depth or shorter. Repeating the calculation for the WIM and HIM, with assumed typical electron densities of 0.2 and 0.005 cm$^{-3}$ respectively and the same magnetic field strength gives 0.32 and 0.008 \radu \ pc\inv \ of Faraday rotation. This, in turn, gives 7 and 140 pc as the depth scales where a Faraday slab would begin to be resolved out in LOFAR observations, for the WIM and HIM respectively. Regions thicker than these values will be significantly depolarized at LOFAR frequencies, while regions thinner then these values will not suffer from significant internal depth depolarization. It should be noted that these depths are based on the assumed parallel magnetic field strength and thermal electron density, and so represent typical order-of-magnitude scales for this behaviour; variations in these parameters will change the required depth.

\subsection{Rejecting Faraday thick models}\label{sec:slabmodel}
An obvious starting point for a physical model of the IC342 field is a Faraday slab model, since most of the field shows two clear emission peaks in the Faraday depth spectrum, which can be interpreted as the signature of a Faraday slab. In this model, the Faraday depth offset from zero (and the variations in this offset with position) would be caused by a foreground Faraday-rotating volume with very little emission, while the emission and the separation between the two features is supplied by a Faraday slab.

It is not possible to determine from the data which emission feature is the leading (nearer to the telescope) edge and which is the trailing. If we assume the brighter feature at lower Faraday depths (top panels of Fig \ref{fig:gradients}) is the leading edge, the foreground Faraday rotation must contribute between $-7$ and +3 \radu \ in front of the slab, and the slab has a thickness of approximately +8 \radu. If we instead assume the weaker feature (bottom panels of Fig \ref{fig:gradients}) is the leading edge, the foreground Faraday-rotating region must contribute +1.5 to +11 \radu \ and the slab has a thickness of approximately $-8$ \radu \ (the negative sign signifying that the Faraday depth decreases with increasing distance). For both cases, there would also be a second Faraday-rotating screen behind the slab, providing negative Faraday-rotation to the background extragalactic sources (Table \ref{table:pointsources}).

For an idealized Faraday slab, both features would have the same intensity, whereas we observe a significant difference, approximately a factor of 2--3 in polarized intensity, between the first and second emission features. This can be explained as a departure from the ideal tophat spectrum, with either a peak in the brighter side of the slab (such as is seen in Fig. 2 of \citealt{Beck12}) or a more gentle decrease in the other side (producing additional depolarization, resulting in a weaker peak in the observed spectrum). If we assume that one of these peaks represents the observed intensity of an idealized Faraday slab and divide by the expected ratio of observed to true amplitude (12\%, as per the previous section), this gives a prediction of the true polarized intensity of the slab. Using 30 K RMSF\inv \ and 10 K RMSF\inv \ for the typical polarized brightnesses of the first and second emission features respectively, this gives intrinsic polarized amplitudes\footnote{To convert from RMSF\inv \ to (\radu)\inv, we use the same method used for the conversion from mJy PSF\inv \ to brightness temperature, adapted to one dimension, and use the fitted Gaussian for RM-CLEAN. The resulting conversion is 0.93 \radu \ RMSF\inv.}
 of 250 or 83 K (\radu)\inv. The two features are separated by approximately 8 \radu, which would mean an intrinsic polarized flux of 2000 to 660 K. If the emission is more complex or turbulent than a uniform Faraday slab, which is almost certainly the case, then the emission will be more strongly depolarized and the intrinsic polarized flux must be higher than these values.

This diffuse emission is not seen in total intensity as it is smooth on the angular scales probed by LOFAR and is correspondingly filtered out. However, the total intensity flux is known from earlier single-dish measurements. The 150 MHz all-sky map from \citet{Landecker1970} shows that the brightness temperature varies across this field from 460 to 600 K. Accounting for the fact that the maximum possible fractional polarization for Galactic synchrotron emission is about 75\% \citep{Rybicki85}, this puts the upper limit for polarized flux at 345 to 450 K. This upper limit requires that the magnetic field be perfectly ordered throughout the emitting volume. For a more realistic combination of turbulent and ordered magnetic fields, this limit drops further.

Since the polarized flux required for this model (660 K or more) significantly exceeds the maximum possible polarized flux consistent with the total intensity (450 K or less), we conclude that our observations cannot be explained by a single Faraday slab or similar feature. This is also supported by the significant differences in morphology between the two observed features. Therefore, a multiple component model is required to explain the observations.

\subsection{A six-component physical model}\label{sec:physicalmodel}
Having rejected the Faraday slab model, we propose a more complicated but physically motivated model, which contains two neutral regions producing the observed Faraday-thin features, three (presumably fully) ionized regions that are Faraday-thick and therefore depolarized and not observed but contribute to the Faraday rotation of the observed features, and the hot ionized Local Bubble.

The Local Bubble is the volume of HIM surrounding the Sun. The estimated depth of the Local Bubble in this direction is 90 pc \citep{Lallement14}, and it is known to have a low electron density of 0.005 cm\inv \ \citep{NE2001}. Again using a typical magnetic field strength of 2 $\upmu$G, the predicted Faraday rotation is 0.7 \radu. Therefore, we do not expect the Local Bubble to contribute significant Faraday rotation of background polarized emission, and the synchrotron emission produced inside the Local Bubble should create a Faraday-thin feature in the Faraday spectrum at a Faraday depth at 0 \radu. The bright emission feature passes through 0 \radu, but it also covers Faraday depths from $-6$ \radu \ to +1 \radu. This indicates the presence of a Faraday rotating screen in front of the emission, so the Local Bubble cannot be the source of this emission feature. Instead, the Local Bubble emission we expect at 0 \radu \ must be fainter than, and thus blended into, the brighter emission feature.

The emission features must be Faraday-thin, to be consistent with the flux calculations in the previous section, and behind at least one Faraday-rotating screen, which must provide the Faraday rotation observed in both components. Below we will identify possible physical causes for the emission features. Shocks from supernova remnants cannot explain our observations because no supernovae remnants are catalogued in the direction of our observations. The available data do not allow us to exclude Faraday caustics or other magnetic phenomena as possible explanations.

Identifying and localizing WNM or HIM volumes of interstellar space is difficult, as there are very few reliable tracers of these phases that are also distance resolved. The \ion{H}{I} 21-cm line traces neutral gas and has been mapped extensively in the Galaxy, but does not provide good distance resolution within the nearest few hundred parsecs. Hot gas can be traced by soft X-ray emission, but this gives no distance information; bubbles of HIM in the Galactic disk are typically identified as voids in the warm medium and by the presence of neutral walls around such bubbles. \ion{Na}{I} absorption of starlight has been used to trace neutral clouds, but comprehensive maps only exist out to a few hundred parsecs \citep[e.g.,][]{Vergely10}. Similar maps of the local ISM have been made using optical extinction and reddening \citep[e.g.,][]{Lallement14, Green15}, which correlate well with the maps of neutral clouds and show the presence of the Local Bubble as a low-density region.

\begin{figure}[h]
\resizebox{\hsize}{!}{\includegraphics{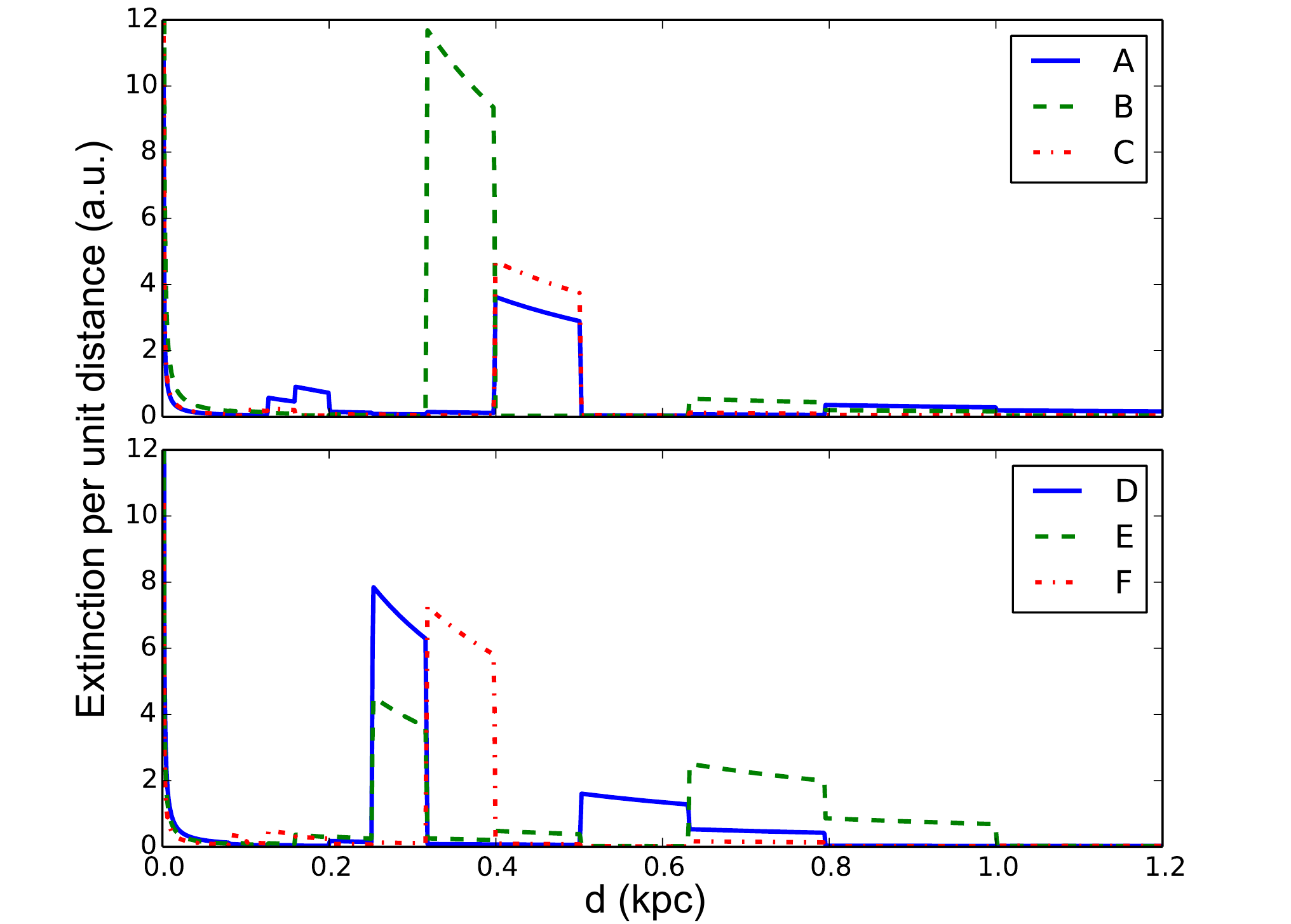}}
\caption{Profiles of the local reddening per unit distance for selected lines of sight in the IC342 field, calculated from the \citet{Green15} reddening model with MWDUST. The labels correspond to the lines of sight shown in Fig. \ref{fig:spectra_1}. All six profiles show the presence of a dust cloud between 200 and 500 pc, and D and E show the presence of a second cloud between 500 and 800 pc.}
\label{fig:dust_profiles}
\end{figure}

\begin{figure}[h]
\resizebox{\hsize}{!}{\includegraphics{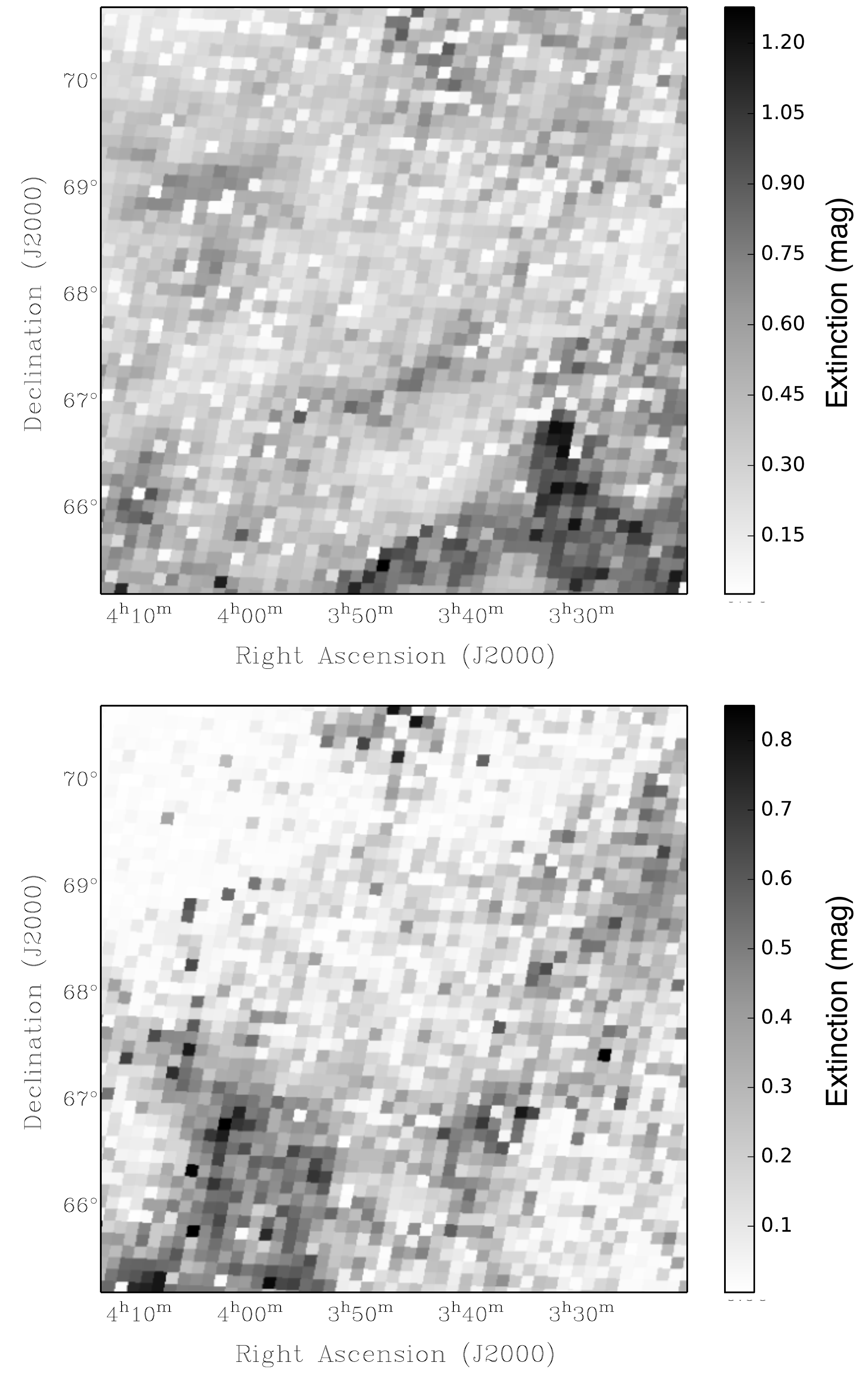}}
\caption{Maps of the reddening caused between 0 and 500 pc (top), and 500 and 1000 pc (bottom), calculated by integrating profiles from Fig. \ref{fig:dust_profiles} over the selected distance range. The pixel size is set by the resolution of the \citet{Green15} model. The top plot shows the presence of a field-filling dust cloud (assumed to be a neutral region) while the bottom shows the presence of a more distant cloud that occupies only part of the field.}
\label{fig:dust_maps}
\end{figure}

We used the software package MWDUST\footnote{https://github.com/jobovy/mwdust} \citep{Bovy16} to probe the dust distribution predicted by the \citet{Green15} reddening model in the IC342 field, as a proxy for neutral clouds in the ISM. This code gives the total reddening to a given position; to determine the position of the dust/neutral clouds, the numerical derivative was taken with respect to distance to give the local reddening per unit distance as a function of distance. The results, for selected lines of sight, are shown in Fig. \ref{fig:dust_profiles}. For all of the lines of sight, there is a clear concentration of dust between 200 and 500 pc (depending on the line of sight), indicating the presence of a cloud that fills the field of view. For lines of sight D and E, which cover the center and lower-left of the field where the fainter emission feature is observed, a second cloud is present between 500 and 800 pc. Based on these profiles, we divided the line of sight into two regions: from 0 to 500 pc, and from 500 to 1000 pc, such that each region contains one distinct region of high local reddening; the resulting maps are shown in Fig. \ref{fig:dust_maps}. The nearer cloud fills the field of view, while the more distant cloud is concentrated along a broad region from the bottom left corner towards the top right. The presence of this nearer cloud is also supported by models of the local ISM \citep{Vergely10, Lallement14}, which generally do not extend far enough in distance to include the second cloud. The model by \citet{Lallement14} shows no bubbles of HIM beyond the Local Bubble in the direction of our data, out to a distance of 500 pc. Therefore HIM regions cannot explain the polarization features we observe.

\begin{figure*}[htb]
\includegraphics[width=17cm]{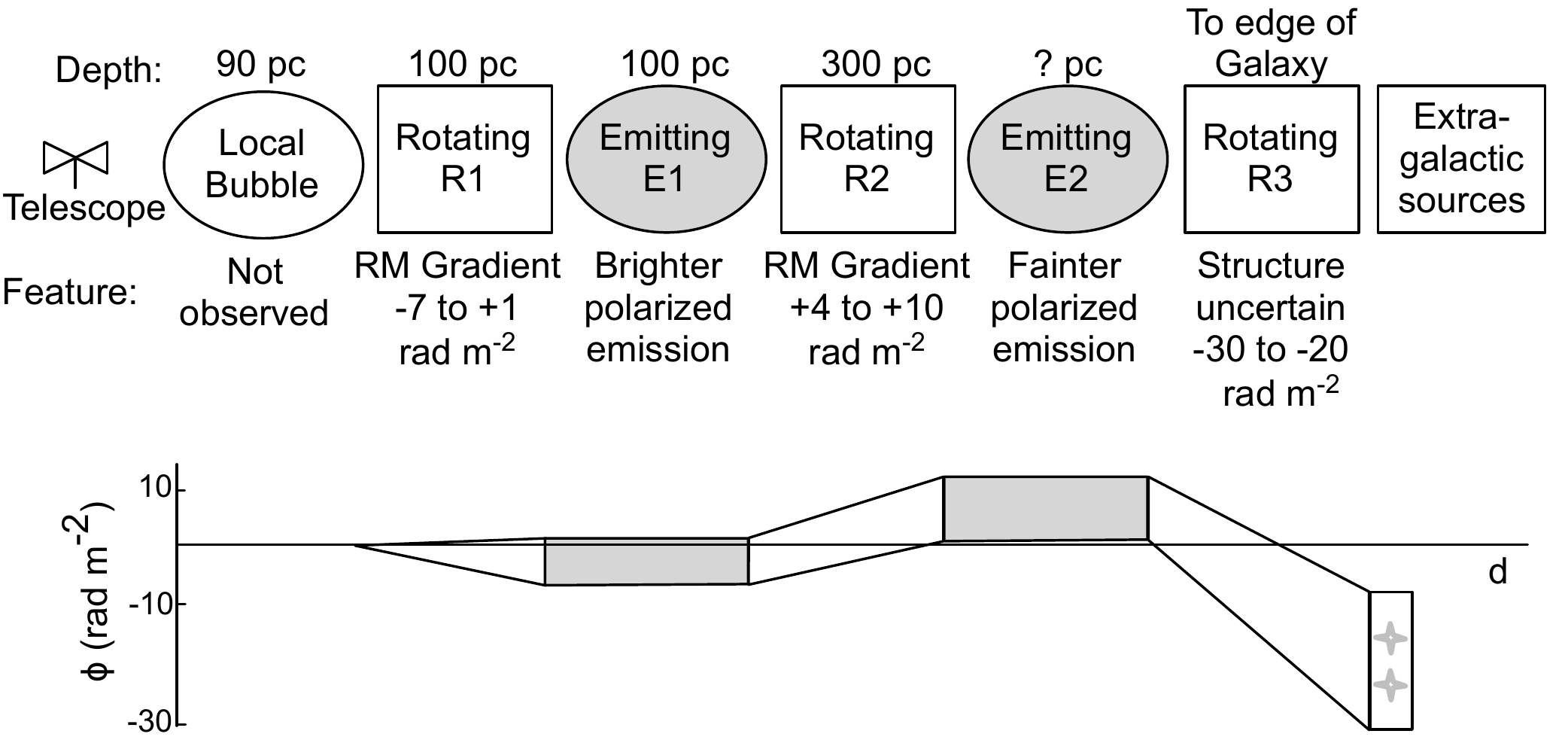}
\caption{A schematic view of the physical model for the IC342 field. The three Faraday-rotating and two emitting regions are labelled, for clarity in the text, and the defining feature of each region is given. The Local Bubble is included as it is a known ISM feature that occupies a portion of the line of sight, but does not contribute significant Faraday rotation and no associated polarized emission is observed. The Faraday depth of each region is shown below, with the two horizontal lines bounding the range of Faraday depths for different positions in the field; for any given direction and distance only one value in that range is appropriate. Grey shading and stars represent diffuse and point-source polarized emission respectively.}
\label{fig:model}
\end{figure*}

Due to the morphological correspondence, we interpret the two emission features in our observations as emission produced in these two neutral clouds, and use the estimated distances and sizes of these clouds to produce a model for the emission and Faraday rotation.  Drawing from the dust models, we begin our model with two warm neutral clouds, the first at a distance of 200 pc, and the second at a distance between 500 and 800 pc, which produce the observed polarized emission. The distance between the Local Bubble and the first neutral cloud we model as a warm ionized region, which provides the observed Faraday rotation of the emission from the first cloud. Between the two clouds is another ionized region which provides the Faraday rotation difference between the two emission features, and beyond the second cloud is some unconstrained volume of ionized gas to the edge of the Galaxy which provides the difference in Faraday rotation between the diffuse emission and the background polarized sources. Fig. \ref{fig:model} gives a schematic view of this model, where the two emitting regions are matched to the two neutral clouds.

The observed emission features were assigned to the neutral clouds based on their morphology. The more distant dust feature runs through the field from the bottom left to the upper right, occupying a very similar part of the field as the fainter emission feature. The nearer dust cloud fills the field of view, as does the brighter emission feature. Based on these similarities, we assigned the brighter feature to the nearer cloud (`E1' in Fig. \ref{fig:model}) and the fainter feature to the more distant cloud (`E2'). The depth of the nearer cloud was estimated from the \citet{Vergely10} and \citet{Lallement14} models to be roughly 100 pc, but the more distant cloud is outside of the region of these models. The \citet{Green15} dust model does not have the distance resolution to estimate the depths of the clouds. We note that the 100 pc depth of `E1' region is greater than the depolarization depth scale discussed at the end of Sect. \ref{sec:theory}, but this can be due to the parallel magnetic field or thermal electron density being slightly lower than the values assumed in that calculation.

At a reference frequency of 150 MHz (near the center of the band for these observations), a typical value for the (total intensity) synchrotron emissivity is about 140 K kpc$^{-1}$ \citep{MarcoFan,Nord06}. The polarized fraction is difficult to estimate, as it depends strongly on position (e.g., the Galactic plane is strongly depolarized) and resolution (coarser observations have increased beam depolarization). The maximum possible polarization fraction of synchrotron emission is about 75\% for Galactic synchrotron \citep{Rybicki85}. An approximate value can be made using Equation 10 of \citet{Burn66}, which states that the polarized fraction is modified from the maximum by the ratio of the energy of the large-scale field to the energy of the total magnetic field when the scale of the random field is smaller than the resolution. Using typical values of 2 $\upmu$G for the ordered field and about 6 $\upmu$G for the total field gives an estimate of about 8\%, but this may be a lower limit as the ordered field estimate is for much larger scales than the expected spatial resolution of our observations (our resolution at 200 pc gives a scale of about 0.3 pc, so all scales above a few pc may be considered as part of the `ordered' field for our purposes). Using this 8\% value, the expected polarized emissivity is 11 K kpc$^{-1}$.

Using the value and the estimated depth of the first emitting region, the predicted polarized synchrotron brightness is 1.1 K, well below the observed value of 30 K. This implies that the perpendicular magnetic field may be significantly stronger than average, or that the magnetic field is more ordered (on the physical scales being probed) than the rough estimates used above. Without a depth estimate for the second cloud, it is not possible to give a predicted polarized brightness.

The first Faraday-rotating region (labelled `R1' in Fig. \ref{fig:model}) is associated with Faraday rotation by the ionized gas between the Local Bubble and the first neutral cloud. As with the Faraday slab model, the first region must provide the spatial gradients in Faraday depth that are observed in both emission features, and must provide all the Faraday rotation present in the first emitting region, corresponding to Faraday depths between $-7$ and +1 \radu. Assuming a path length of approximately 100 pc, and a thermal electron density of 0.1 cm$^{-3}$, the required average parallel magnetic field strength ranges from $-$0.86 to +0.12 $\upmu$G, with the magnetic field directed away from the Earth (negative) in the lower left corner and towards the Earth (positive) throughout the rest of the field.

The second Faraday-rotating region (`R2') provides the Faraday depth offset between the two diffuse emission features, which varies with location between +4 and +10 \radu. Assuming a depth of about 300 pc for this inter-cloud region, the average product of the electron density and parallel magnetic field needed to produce this Faraday rotation is between 0.016 and 0.042 $\upmu$G cm$^{-3}$. If we again assume a thermal electron density of 0.1 cm$^{-3}$, this gives an average parallel magnetic field strength of +0.16 to +0.42 $ \upmu$G.

The final Faraday-rotating region (`R3') represents all Faraday rotation between the second cloud and the extragalactic polarized sources. \citet{Oppermann14} used published extragalactic Faraday rotation measurements to produce an all-sky map of the Galactic foreground contribution. For our field, their map gives Faraday depths ranging from $-54$ \radu \ to +16 \radu, with a typical error of 10 to 30 \radu. Since this error is much larger than the range of Faraday depths we observe, we concluded it was not meaningful to produce a difference map between the \citet{Oppermann14} map and the second emission feature, which would represent the Faraday rotation in region `R3'.

\section{Discussion}\label{sec:discussion}
The largest discrepancy between the models and the observations is the intensity of the polarized synchrotron emission. This is not a surprising result given that this field is in the Fan region, which has been long known to have abnormally high polarization \citep{Brouw76}. This is further supported by LOFAR observations of other regions of the sky, which have observed polarized brightnesses between 1 and 15 K \citep{Jelic14,Jelic15}, and observations with the Murchison Widefield Array, which have observed an average polarized brightness of 4 K at 154 MHz \citep{Lenc16}. Given the unusually high polarization of the Fan region, an enhancement in the perpendicular magnetic field or the degree of order in the magnetic field would be quite reasonable.

Our model presented in Sect. \ref{sec:physicalmodel} assumes that the Faraday thin emission comes from mostly neutral regions associated with the warm neutral phase of the ISM. It could be possible that one or both emission features correspond to a Faraday caustic, particularly the fainter feature as that emission has the most positive Faraday depths and could represent the transition from parallel fields oriented towards the Sun (producing positive Faraday rotation) to fields oriented away from the Sun (producing negative Faraday rotation).  This alternative is effectively indistinguishable from the two neutral cloud model of the previous section, but would require a substantial path length with very small parallel magnetic fields in order to produce enough polarized intensity at the same Faraday depth. The more distant neutral cloud in the model is less certain to exist than the nearer, as it is beyond the range of the \citet{Lallement14} model and the morphological correspondence between the \citet{Green15} extinction map (bottom of Fig \ref{fig:dust_maps}) and the fainter emission feature (bottom left of Fig \ref{fig:gradients}) is weak. This feature could also be explained by a  bubble of HIM without affecting the model significantly. Another possibility, as discussed in Sect. \ref{sec:theory}, is enhanced magnetic fields from a shock or compression. There are no known supernova remnants or other features that might indicate such a shock, so we did not consider this for our model.

It is important to note that in this model it is not that the synchrotron emission or intrinsic polarized fraction is enhanced in the neutral regions, compared to the ionized regions, but rather that these are the only portions of the line of sight that are not strongly depolarized at low frequencies. The magnetic field can have identical properties between the neutral and ionized regions, without affecting this model. The transition between a strongly Faraday-rotating ionized medium and a weakly Faraday-rotating (mostly-) neutral region, combined with the limited physical depth of the neutral regions, produces a very narrow feature in the Faraday spectrum that does not depolarize much compared to the other polarized emission along the line of sight.

In this model we have considered only the effects of depth depolarization, and not beam depolarization. Depth depolarization causes a Faraday-thick emitting and rotating region to depolarize, but does not affect the polarized intensity of background emission (unless the background emission overlaps in Faraday depth, as might occur if the parallel component of the magnetic field reverses sign along the line of sight). Beam depolarization, which can be produced by unresolved gradients in Faraday depth, will cause some depolarization of background emission passing through a Faraday rotating foreground \citep{Tribble91,Sokoloff98, Schnitzeler2015b}. This is most likely present in our observations, causing the observed polarized intensities to be lower than the true values. Since the values of polarized intensity were not important for our analysis (beyond the observation that they are already quite high, even without accounting for beam depolarization), we did not include any beam depolarization in our modelling.

If we assume that the observed emission features are Faraday thin and not significantly depolarized, the integrated polarized intensity of each diffuse polarized feature (Fig.~\ref{fig:gradients}, left panels) should represent the intrinsic polarization of the emitting regions. The variations in the polarized intensity with position on the sky may reflect variations in the local synchrotron emissivity which are caused by variations in the perpendicular magnetic field. A detailed analysis of the properties of the integrated intensity may yield interesting measurements of the properties of the emitting region, but such analysis is beyond the scope of this paper.

Our model also assumes that the Faraday rotation of the first emission feature was caused by an ionized region outside the Local Bubble when estimating the magnetic field strength. It is also possible that the Local Bubble wall may provide a significant amount of this Faraday rotation, if it has enhanced magnetic field strength and free electron density \citep[e.g., such as observed in the W4 superbubble by][]{Gao15}. It is also possible that part of the variation in Faraday depths across each emission feature is caused by changing path lengths through the ionized regions, if the distances to the clouds vary significantly between different positions in the field.

The morphological correspondence between the dust maps and the observed polarized emission was used to motivate the presence of the neutral clouds in the model, but the correlation between regions with high dust density and high polarized emission is actually quite poor. We can explain this imperfect correspondence because the polarized synchrotron intensity depends on the path length of neutral (or low-ionization) material, and not the column density. For regions of higher dust column density, it is not possible to distinguish between lines of sight with long path lengths of lower density neutral material or shorter lengths of higher density material. For our analysis it is not important what quantity of dust present, but instead where it is present in sufficient quantity to serve as an indicator of the neutral phase of the ISM.

Further evidence that the fainter emission is likely to be more distant is in the characteristic angular scale of the emission. From a visual inspection of Fig. \ref{fig:gradients}, it appears that the brighter feature has more emission on larger angular scales (the long, mostly straight depolarization canals are a clear signature of this), while the fainter emission clearly has much more structure on smaller scales. If we assume that the characteristic angular scale is caused by the characteristic turbulent length scale in the emitting volume and that this scale is approximately the same for both features, then the fainter emission must be more distant. If we assume that the structure comes from depolarization effects in Faraday-rotating regions in front of the emission, and that these depolarization effects are tied to the turbulent length scale in the Faraday-rotating regions, then the same argument holds and the fainter emission must be more distant. A quantitative analysis of the characteristic scales and angular power spectra is beyond the scope of this paper, but should be investigated in follow-up studies.

This type of modelling can benefit significantly from the inclusion of rotation measure and dispersion measure data (which measure the column density of free electrons) from pulsars with independent distance estimates.  Of the 17  pulsars listed in the ATNF catalog\footnote{http://www.atnf.csiro.au/people/pulsar/psrcat/} within 10 degrees of IC342, 16 have DMs, 4 have RMs, but only two have independent distance measurements. Both of these are beyond 2 kpc, well outside the distance range of our model, so they are not useful for constraining either the Faraday rotation or the electron density. We did not include these pulsars in our analysis, but future modelling on other fields should consider pulsar measurements.

\section{Summary and Conclusions}
\label{sec:conclusions}
We have observed a 5$^\circ$ by 5$^\circ$ region centred on the nearby galaxy IC342 using LOFAR in the frequency range 115--178 MHz, and performed Faraday tomography to detect the foreground Galactic diffuse polarized synchrotron emission. We clearly detect two emission features, overlapping in position but separated in Faraday depth. Both features are distributed in Faraday depth with similar gradients, but with very different morphologies in integrated intensity.

We have performed simulations showing the extent of the depolarization of Faraday-thick structures at LOFAR frequencies. Faraday slabs, which are defined by a tophat function in the Faraday profile and represent regions of uniform emission and Faraday rotation, are strongly depolarized: they retain only 12\% of their true amplitude at the edges, producing the appearance of two low-intensity Faraday-thin peaks. Smoother features in the Faraday profile would be more strongly depolarized.

From the strong depolarization shown in these simulations, and a comparison of the observed polarized intensity compared to the total intensity, we argue that these features cannot be the edges of a Faraday slab or other Faraday thick structure, and represent two Faraday thin emission regions. Such emission regions require a volume without significant Faraday rotation, so we further argue that these emission regions probably correspond to mostly neutral clouds within the nearby ISM large enough to produce significant synchrotron emission. We have inspected reconstructed maps of the ISM, and found there is evidence for two neutral clouds along the lines of sight we observed. Using the estimated sizes and distances to these neutral clouds, we proposed a model where these two neutral regions produce the Faraday-thin polarized emission, while (depolarized) ionized regions through the remainder of the line of sight provide the observed Faraday rotation structure.
Using estimated sizes and distances to these clouds, we have modelled the synchrotron emission and Faraday rotation for lines of sight through this region. We find that even in the Faraday-thin case, where there is no depth depolarization present, we observed much more polarized intensity than can be explained using typical values for relevant parameters. This is not surprising, as our field is in the Fan region, which is known for anomalously high polarization. 
We estimated that the strength of the parallel magnetic field required to produce the observed foreground Faraday rotation is  $-$0.86 to +0.12 $\upmu$G (where positive is orientated towards the Earth, negative away from the Earth).

To confirm that the observed emission features are tied to these neutral clouds, similar observations over a large area of the sky would be very useful. These would allow for the large-scale morphology of the emission to be observed and correlated against the boundaries of the neutral clouds inferred from extinction and \ion{Na}{I} absorption. Such observations would be best done at low Galactic latitudes, where the locations of neutral clouds are best constrained by the ISM models.

If confirmed, this provides us with a very powerful method to map out the magnetic field (parallel to the line of sight) inside the local ISM. There are many known neutral clouds within 500 pc of the Sun, which could be used to produce models of the magnetic field in the local ISM in the same way that the observed RMs of pulsars and extragalactic sources are used to model the large-scale field of the entire Galaxy.

This method relies on the properties of depolarization at very low frequencies. Emission features that are extended in Faraday depth are very strongly depolarized at low frequencies, meaning that they can be effectively filtered out based on the choice of observing frequency, leaving only Faraday-thin components that can be isolated and studied individually. This makes low-frequency Faraday tomography a unique way to probe the magnetism of our Galaxy.

\begin{acknowledgements}
We thank the referee for their feedback, which has significantly improved the manuscript. This work is part of the research programme 639.042.915, which is (partly) financed by the Netherlands Organisation for Scientific Research (NWO). AMMS and DDM gratefully acknowledge support from ERCStG 307215 (LODESTONE). \\

LOFAR, the Low Frequency Array designed and constructed by ASTRON, has facilities in several countries, that are owned by various parties (each with their own funding sources), and that are collectively operated by the International LOFAR Telescope (ILT) foundation under a joint scientific policy.
\end{acknowledgements}

\bibliographystyle{aa} 
\bibliography{References} 

\appendix
\section{Simulating Faraday slabs} \label{app:slabs}
\begin{figure}[h]
\resizebox{\hsize}{!}{\includegraphics{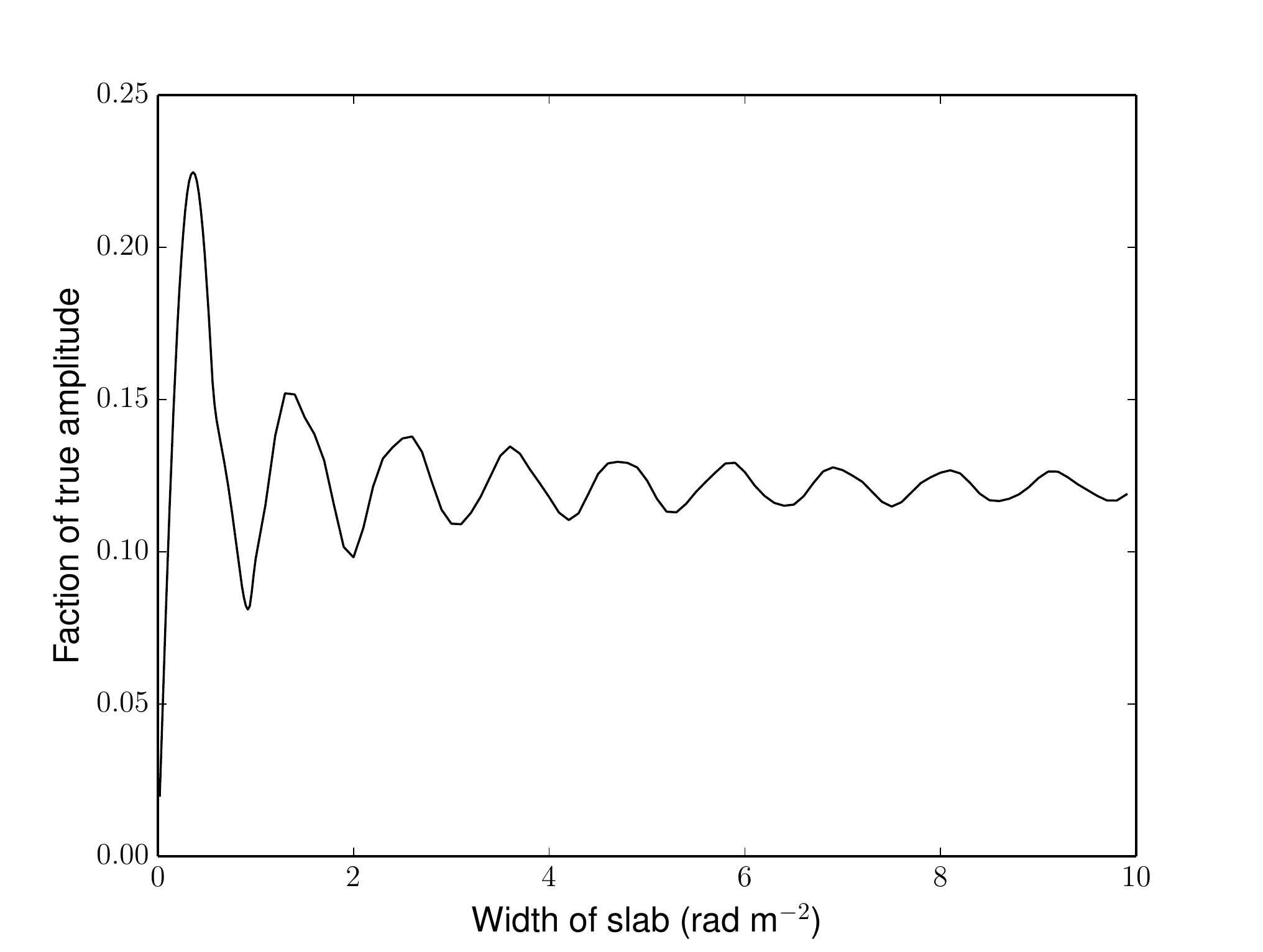}}
\caption{Measured amplitude of ideal Faraday slabs, for the same bandwidth as the IC342 observations, as a function of the width of the slab. For widths significantly wider than the RMSF (1 \radu), the amplitude oscillates around 12\% of the intrinsic amplitude, with variations of about 1\%. }
\label{fig:slab_amplitude}
\end{figure}

To determine the degree of depolarization that could be expected in LOFAR observations, we performed simulations of Faraday slabs of different widths but fixed amplitude (where we define the amplitude as the magnitude of the Faraday profile, which has units of spectral flux density (or brightness temperature) per unit Faraday depth), using the same frequency coverage as the IC342 observations, and measured the resulting simulated peaks in the Faraday spectrum. The resulting `measured' amplitude of the peaks is shown in Fig. \ref{fig:slab_amplitude} as a function of the width of the slab. For widths greater than about 2 \radu \ (twice the width of the RMSF), the amplitude varies between 11 and 13\% of the true amplitude.

The weak dependence on the width is a consequence of the way we have defined the Faraday slab, with a fixed Faraday spectrum amplitude. This results in the intrinsic polarization (the hypothetical polarization at $\lambda^2 = 0$) being equal to the product of the amplitude, $A_\phi$, and the width of the slab, $\Delta \phi$. The polarized intensity as a function of wavelength is then defined as $P(\lambda^2) = A \Delta \phi \frac{|\sin(\Delta \phi \, \lambda^2)|}{\Delta \phi \, \lambda^2}$. So for a fixed bandwidth, the only effect of changing the width is the number of oscillations of the sine term, which only weakly affects the observed amplitude. Therefore, the 11 to 13\% figure given above applies to all Faraday slabs with widths greater than about 2 \radu \ when observed by LOFAR.

\end{document}